# Machine Learning-Based Basil Yield Prediction in IoT-Enabled Indoor Vertical Hydroponic Farms

Bouzid Emna[1], ebouzid@ualr.edu
Baccar Noura[2]*, Noura.baccar@medtech.tn
Iqbal Kamran[1], kxiqbal@ualr.edu
Chaouch Yassine[2], yassine.chaouch@medtech.tn
Ben Youssef Fares[2], fares.benyoussef@medtech.tn
Regayeg Amine[2], amine.regayeg@medtech.tn
Toumi Sarra[3], sarra.toumi@medtech.tn
Nsir Houda[3], houda.nsir@medtech.tn
Mseddi Amina [2], amina.mseddi@medtech.tn
Costelle Leila[3], leila.costelle@medtech.tn

[1] *Electrical and Computer Engineering, University of Arkansas, Little Rock, Arkansas, United States*
[2] *Computer Systems Engineering Department, Mediterranean Institute of Technology, South Mediterranean University, Tunisia*
[3] *Renewable Energy Engineering Department, Mediterranean Institute of Technology, South Mediterranean University, Tunisia*

\* *Corresponding author*

## Abstract

As agriculture faces increasing pressure from water scarcity, especially in regions like Tunisia, innovative, resource-efficient solutions are urgently needed. This work explores the integration of indoor vertical hydroponics with Machine Learning (ML) techniques to optimize basil yield while saving water. This research develops a prediction system that uses different ML models and assesses their performance. The models were systematically trained and tested using data collected from IoT sensors of various environmental parameters like $CO_2$, light... The experimental setup features 21 basil crops and uses Raspberry Pi and Arduino. 10k data points were collected and used to train and evaluate three ML models: Linear Regression (LR), Long Short-Term Memory (LSTM), and Deep Neural Networks (DNN). The comparative analysis of the performance of each model revealed that, while LSTM showed high predictive capability and accuracy of 99%, its execution time was 10 times longer than LR and its RAM usage was about 3 times higher than DNN's when simulated on a standard CPU environment. Conversely, the DNN model had an accuracy rate of 98%. This proves an efficient balance between computational speed and prediction quality, which makes this model well suited for real-life deployment. Moreover, LR excelled in fast processing of basic prediction with an execution time of 11 seconds. This makes the LR model more suitable for low-complexity or resource-limited applications. These performance trade-offs highlight the potential of

DNN-based solutions for building responsive, high-accuracy decision-support systems tailored to agricultural environments, making it suitable for future edge-device deployment.

## 1. Introduction

In low- and middle-income countries, such as Tunisia, water scarcity presents a pressing challenge. Tunisia is categorized as a country facing high water stress [1]. The country's renewable water resources are below the water scarcity threshold of 1,000 cubic meters per capita per year. Notably, high water consumption in agriculture exacerbates this issue. In fact, agriculture in Tunisia accounts for around 80% of total water consumption, which is far above the global average [2]. Water scarcity would result in a significant decrease in agricultural productivity [3].

Some solutions already exist to address this issue, like enforcing the laws regarding water usage and rehabilitating existing water infrastructure to enhance system performance [4]. However, a promising solution lies in Hydroponics. This innovative farming technique has the potential to mitigate water scarcity problems by drastically reducing water consumption while simultaneously enhancing agricultural productivity [5] [6]. By fostering a more efficient and sustainable approach to cultivation, Hydroponics offers a pathway to address the critical issue of water scarcity.

By utilizing vertical space and circulating water-based nutrient solutions, this technique highly reduces water consumption compared to traditional soil-based farming methods. Moreover, the state of the art reflects a growing interest in the integration of IoT and automation technologies into hydroponic agriculture [7]. Many studies focus on developing systems for monitoring and controlling environmental parameters critical to plant growth, such as pH, conductivity, temperature, and humidity [8] [9] [10]. While these studies lay the groundwork for precision agriculture in hydroponics by creating more efficient and controlled growth environments, they often do not explicitly address the scalability of their systems or the comprehensive use of data analytics to optimize water and nutrient usage. Building on this base, our research introduces comprehensive data analytics designed for vertical hydroponic systems, which use IoT to optimize water and nutrient consumption

Vertical hydroponics is another promising approach that addresses water scarcity by efficiently using space and resources.

In alignment with these objectives, we formulated a set of research questions to structure and guide our investigation, as outlined below:

**Research Question:** How can Machine Learning Algorithms be used with IoT data of Hydroponics Vertical Farming to predict future data points, specifically, forecasting basil crop growth?
- **Sub-question 1:** What are the critical data points required to develop an analytics model for plant-growth prediction in vertical hydroponic systems of green leaf plants, specifically basil?
- **Sub-question 2:** How can IoT data be fit for Machine Learning?
- **Sub-question 3:** How do different predictive Machine Learning models, specifically Linear Regression, LSTM and DNN, compare in performance, accuracy, and resource-usage, when implemented with hydroponic system data?

This paper is organized as follows. Section 2 presents a comprehensive literature review focusing on the application of machine learning techniques in indoor agriculture. Section 3 outlines the materials and methods employed in this study, followed by Section 4, which details the data collection process. Section 5 describes the machine learning models utilized and the features extracted for analysis. In Section 6, we discuss the performance assessment metrics and the prediction of plant growth patterns. Section 7 presents the experimental results alongside a comparative analysis of the models. Finally, Section 8 concludes the paper and suggests directions for future research.

## 2. Literature Review on Machine Learning in Indoor Agriculture

Recent advancements in smart agriculture have demonstrated the potential of Machine Learning (ML) to support environmental monitoring, predictive analysis, and decision-making in indoor controlled farming systems. Despite the growing body of research, many projects fail to elaborate on the ML models used, the features selected for training, or the interpretability of the outcomes. For instance, in the work by Mduma and Mayo (2024) [11], ML was used to detect maize diseases in a Tanzanian use case, yet the specific algorithms and feature engineering processes remain undocumented. Similarly, while [12] discusses the benefits of remote data access through smart farming, it lacks technical insights into how the data is modeled or analyzed. Other studies [13], [14], [15] describe the deployment of IoT sensors for environmental monitoring; capturing variables such as temperature, humidity, and $CO_2$; but omit the predictive models applied to these datasets.

### 2.1. Linear Regression

To address such gaps, the application of regression-based models offers a fundamental starting point for analysis and forecasting. Linear regression, in particular, serves as a baseline model to identify correlations between sensor readings and target variables, such as light intensity or $CO_2$ levels. Its variants, such as Ridge, Lasso, and Bayesian regression, further enhance performance by handling multicollinearity and incorporating prior domain knowledge [16–18].

In the environmental sciences, linear regression has been widely adopted for trend analysis and forecasting. For example, M. Leon et al. [19] employed it to examine the relationship between temperature and carbon emissions, demonstrating how even simple models can yield actionable insights. This motivates our exploration of linear regression techniques for modeling sensor-derived environmental data in smart indoor agriculture systems.

Research done by Harsapranata et al [20], is a forecast of the nutritional needs to support plant growth, based on Linear Regression analysis. Their algorithm is influenced by the pH of the nutrients, the height of the nutrient liquid in the reservoir, and the dissolved nutrient particles in the reservoir. They worked with the paprika plant and the calculations were made every 4 days, within 56 days.

Multiple linear Regression was used in the control of hydroponic acidity (pH) in a work done by H. Helmy et al [21], where the goal was to maintain the pH value at a stable level. This research discussed the details of an automated pH control system by using multiple linear regression. The accuracy result of the multiple linear regression tests with several reservoirs and several set points was 94.84%. For the overall system accuracy test, the result was 89.37%.

## 2.2. Long- Short Term Memory (LSTM)

While traditional regression models offer interpretability and are suitable for linear relationships, they often fall short when dealing with complex temporal dependencies in dynamic environments such as vertical hydroponic systems. To address this, recent studies have increasingly turned to deep learning approaches, particularly those designed for sequential data.

Among these, Long Short-Term Memory (LSTM) networks have emerged as a powerful tool for modeling time series due to their ability to retain long-range dependencies. LSTM is a specialized type of recurrent neural network (RNN) developed specifically for sequence prediction tasks [22]. Unlike feedforward architectures such as Multi-Layer Perceptrons (MLPs), which are stateless and limited in handling sequential patterns, LSTMs maintain an internal memory that allows them to capture the temporal structure of input data.

Multiple LSTM variants have been introduced to enhance performance in specific contexts, including vanilla LSTM, stacked LSTM, CNN-LSTM, encoder-decoder LSTM, bidirectional LSTM, and generative LSTM models. These architectures are especially useful for applications that involve forecasting, anomaly detection, and noise suppression.

Kim, Yu, and O'Hara [23] compared LSTM-based noise filtering with traditional methods such as Kalman and moving average filters. Their findings demonstrated that the LSTM filter achieved superior performance in minimizing false alerts within threshold-based agricultural automation systems. They also noted a lack of prior application of ML-based filters in large-scale aquaponics or agriculture sensor networks.

The research by Suebsombut et al. [24] presents the use of LSTM and Bi-directional LSTM in the prediction of soil moisture. They collected data by using sensors in a greenhouse in Chiang Mai in Thailand. They also used simple imputation and mean to prepare the data and replace the missing values. 14 features and 17,749 samples were extracted and used for the LSTM model implementation. They used 4 dropout layers and 12 dense units as output layers.

Further building on this, Yu and Kim [25] developed an LSTM anomaly detection filter capable of identifying deviations in sensor data streams with high precision. Their approach outperformed statistical filters and, importantly, operated in an unsupervised manner with zero false positives. Although anomaly detection can be prone to false alarms due to unobserved but normal behavior, their LSTM-based method effectively learned baseline patterns to distinguish genuine anomalies from benign variation.

### 2.3. Deep Neural Networks (DNN)

While LSTM networks have demonstrated strong capabilities in handling sequential and time-dependent data, particularly filtering and anomaly detection in agricultural settings, they are only one class within the broader family of deep learning methods. In scenarios where the interactions among numerous environmental variables are highly complex and not strictly temporal, alternative architectures may offer complementary strengths. Deep Neural Networks (DNNs) provide a flexible and powerful framework for capturing non-linear relationships and extracting hierarchical features from large and multidimensional datasets.

Compared to traditional Machine Learning algorithms such as support vector machines, perceptrons, decision trees, and k-nearest neighbors, DNNs have significant advantages in extracting features at different levels of abstraction and thereby learning more complex patterns [26].

The interaction of multiple variables in this type of vertical hydroponic farming project needs sophisticated modeling techniques for prediction, such as DNNs. They are a subset of Machine Learning and are well-suited to handling such complexity due to their ability to model non-linear relationships and learn from large datasets [27], [28], [29].

While promising, DNNs need substantial amounts of data for training and high computational power. The black-box nature of DNNs also presents some interpretability challenges [30]. Future research should focus on creating more efficient and transparent DNN architectures and exploring some unsupervised learning approaches.

### 2.4 Comparative Machine Learning Models in Agriculture

Recent studies have employed a wide range of machine learning and deep learning approaches to improve yield prediction in agriculture. Traditional models such as Linear Regression (LR) provide baseline interpretability but are often unable to capture complex, non-linear crop–environment relationships [31]. Long Short-Term Memory (LSTM) networks, by contrast, excel in modeling temporal dependencies and have been widely adopted in agricultural time series forecasting. Similarly, Deep Neural Networks (DNNs) offer flexible nonlinear approximations that can capture high-dimensional patterns in sensor data.

Beyond regression-based and neural approaches, several other machine learning techniques have demonstrated strong potential in yield prediction and precision agriculture.

Random Forest (RF) has proven to be one of the most reliable ensemble methods for modeling nonlinear relationships and variable interactions in crop yield forecasting. Recent studies, such as Abate et al. [32] and Salami [33], demonstrated that RF provides high accuracy and robustness when handling noisy IoT and satellite data in heterogeneous agricultural environments. However, RF models often face limitations in interpretability and computational efficiency, particularly when implemented on resource-constrained IoT or edge devices [34].

Extreme Gradient Boosting (XGBoost) and other boosting algorithms have also achieved competitive performance in large-scale yield prediction tasks. Ensemble and gradient-boosting frameworks, including hybrid RF-XGBoost approaches [35], frequently outperform conventional regression and decision-tree models in accuracy and generalization. Other works in [36] and [37] reported that XGBoost scales well across spatially distributed datasets but requires meticulous hyperparameter tuning and is prone to overfitting in small or highly correlated datasets.

Autoregressive Integrated Moving Average (ARIMA) remains a strong statistical baseline for univariate yield time-series forecasting, particularly when long-term historical data are available. Studies integrating ARIMA and LSTM frameworks for temporal crop forecasting [38] highlight the complementary strengths of linear and nonlinear modeling. ARIMA efficiently captures temporal autocorrelation, whereas LSTM addresses nonlinear dependencies often present in IoT-based precision-farming data, which is the case of this study.

Meanwhile, Convolutional Neural Networks (CNNs) have been effectively applied in precision agriculture to extract spatial and spectral features from canopy and UAV imagery for yield estimation. CNN-based architectures can automatically learn discriminative spatial patterns related to plant health, canopy density, and stress indicators, enabling spatially explicit yield mapping. As demonstrated by Nevavuori et al. [39], CNNs trained on UAV imagery substantially reduce yield-prediction uncertainty compared with conventional approaches. However, these models require extensive labeled datasets, careful preprocessing, and significant computing for training, which constrains deployment on low-cost or real-time IoT platforms.

Recent research trends increasingly favor hybrid and ensemble approaches that combine statistical, neural, and metaheuristic optimization methods. Fariz et al. [40] integrated bio-inspired optimization with deep learning for improved generalization in yield prediction but reported high computational cost and energy demands. Jovanovic et al. [41] and Alharbi et al. [42] demonstrated that hybrid metaheuristic-tuned architectures can enhance performance yet remain experimental, with limited validation on large-scale or diverse crop datasets. Furthermore, Banerjee et al. [43] emphasized that integrating remote sensing and landscape metrics within ML models enhances spatio-temporal prediction capacity, though their study focused more on crop pattern monitoring than direct yield estimation.

Overall, these approaches collectively advance predictive modeling in precision agriculture. However, their practical deployment requires balancing accuracy, interpretability, computational efficiency, and scalability, particularly for IoT-based smart farming systems where real-time, energy-efficient prediction is essential.

By situating LR, LSTM, and DNN alongside RF, XGBoost, ARIMA, and CNN, our study acknowledges the breadth of available techniques and their relative strengths. While our experimental evaluation focused on LR, LSTM, and DNN, future benchmarking will incorporate ensemble models and image-based deep learning approaches to further improve generalizability and robustness.

In summary, the literature highlights a growing interest in applying Machine Learning techniques to smart agriculture, particularly in controlled environments such as vertical hydroponic systems. As shown by *Table_1* below, while traditional models like Linear Regression and Decision Trees offer interpretability and simplicity for tasks involving limited variables, they fall short in capturing the temporal and non-linear dynamics present in more complex scenarios. Deep Learning models, especially LSTM networks and other recurrent architectures, have shown promise in time series analysis, anomaly detection, and environmental forecasting due to their ability to model sequential dependencies and learn from high-dimensional data. Building upon these insights, we conducted a comparative analysis of various machine learning approaches, evaluating their performance in relation to task-specific requirements such as accuracy, computational efficiency, scalability, and interpretability.

*Table_1* below shows a comparative analysis of Machine Learning models within this context:

*Table_1: Comparative analysis of Machine Learning models within the context of precise agriculture*

| Model / Technique | Main Strengths | Limitations | Representative Applications | Relevance to Indoor / Hydroponic Farming |
|---|---|---|---|---|
| Linear Regression (LR) | Simple, interpretable, low computational cost; effective for identifying linear relationships between environmental features and plant growth variables. | Assumes linearity; sensitive to multicollinearity and outliers; limited performance for complex, non-linear dynamics. | Temperature–$CO_2$ correlation analysis [19]; nutrient-demand estimation for paprika [20]; pH control in hydroponic reservoirs [21]. | Baseline predictive model for understanding dependencies among IoT sensor features (e.g., pH, TDS, light, $CO_2$). |
| Long Short-Term Memory (LSTM) | Captures temporal dependencies; robust to noise; suitable for sequential sensor data; adaptable to multivariate time-series forecasting. | Requires large datasets; higher computational demand; less interpretable than linear models. | Noise filtering and anomaly detection in agriculture [23], [25]; soil-moisture forecasting using Bi-LSTM [24]. | Ideal for sensor-stream forecasting in hydroponic systems, modeling dynamic variables such as humidity, nutrient flow, and temperature over time. |
| Deep Neural Networks (DNN) | Handles high-dimensional, nonlinear data; extracts hierarchical features; generalizes across complex feature spaces. | Requires substantial data and compute; risk of overfitting; interpretability challenges ('black box'). | Environmental variable modeling and complex sensor integration [26–29]. | Effective for multivariate prediction tasks (e.g., yield or growth index) where variables are not strictly temporal. |
| Random Forest (RF) | Ensemble-based; manages non-linearities and feature interactions; resistant to overfitting; good performance on small-to-medium datasets. | High memory use; limited transparency of decision boundaries; slower for very large datasets. | Yield prediction with IoT and remote-sensing data [32], [33]; greenhouse environmental modeling. | Useful for feature importance analysis and medium-scale yield estimation using IoT sensor data. |

| Extreme Gradient Boosting (XGBoost) | High accuracy and scalability; handles missing values well; performs well on tabular data. | Requires careful hyperparameter tuning; prone to overfitting correlated datasets; longer training time. | Multi-crop yield forecasting [35–37]; spatio-temporal modeling of distributed farms. | Suitable for high-resolution sensor fusion and hybrid modeling in precision indoor agriculture. |
|---|---|---|---|---|
| Autoregressive Integrated Moving Average (ARIMA) | Strong baseline for univariate forecasting; interpretable; useful for trend and seasonality detection. | Cannot capture nonlinear or multivariate dependencies; limited for irregular or multi-sensor datasets. | Hybrid ARIMA–LSTM for yield forecasting under climate variation [38]. | Benchmarking model for analyzing trends in time-series sensor-derived parameters. |
| Convolutional Neural Networks (CNN) | Excels in spatial and spectral feature extraction from imagery (UAV, multispectral, canopy). | Requires large labeled datasets; high computational cost; less effective for non-image data. | UAV-based yield prediction and canopy-stress mapping [39]. | Important for visual-based growth monitoring and automated phenotyping in indoor systems. |
| Hybrid / Metaheuristic Models (e.g., ANN + Optimization, RF–XGBoost) | Combine complementary strengths of multiple algorithms; improved generalization and prediction accuracy. | Computationally intensive; complex to interpret and tune; limited validation on large-scale datasets. | Hybrid Coati-optimized ANN [40]; metaheuristic-tuned neural networks [41]; optimization for smart agriculture [42]; landscape-metric ML [43]. | Promising for multi-objective optimization, yield forecasting, and decision support where both interpretability and accuracy are required. |

Based on the comparative overview presented in Table 1, it is evident that no single Machine Learning algorithm universally outperforms others across all agricultural prediction tasks. Rather, the choice of model depends on the characteristics of the dataset, the complexity of the relationships among features, and the specific objectives of the vertical farming system. Linear Regression and other simple models remain valuable for exploratory analysis and low-dimensional problems due to their interpretability and minimal computational requirements. However, as the volume, dimensionality, and temporal dependency of IoT data increase, more advanced architectures such as LSTM and Deep Neural Networks (DNNs) become essential for capturing dynamic, non-linear interactions among environmental variables. Ensemble methods like Random Forests and XGBoost provide balanced performance where interpretability and accuracy are both important, while hybrid and metaheuristic approaches offer promising extensions for optimizing predictive accuracy in complex, data-rich environments.

Consequently, this study focuses on Linear Regression, LSTM, and DNN as representative models capable of addressing different aspects of environmental data forecasting in vertical hydroponic systems, evaluated through robust statistical metrics including MAE, RMSE, %SEP, and $R^2$ to ensure a comprehensive assessment of prediction reliability and precision.

## 3. Materials and Methods

### 3.1. Experimental Design Research Methodology

This research consists of studying and building artifacts using the Experimental Design research method. Experimental Design is a technique that enables scientists and engineers to efficiently assess the effect of multiple inputs, or factors, on measures of performance, or responses. Compared to one-factor-at-a-time, trial-and-error approaches, a well-designed experiment can provide clear results while dramatically reducing the required amount of testing [44].

For this work, the research started by identifying the specific problem related to the Tunisian climate and economy as well as the current developments in Technology with farming. This combination of problem and opportunity serves as the driving force behind this work, guiding the design and development of our innovative solution.

The main activity in this research is the creation and the testing of the artifact, that is composed of a physical farming unit collecting data through sensors and a software side that analyzes the data and prepares it for Machine Learning predictive models. The experimentation was done at the Mediterranean Institute of Technology (MedTech) site [45], while the analysis and modeling were performed at the University of Arkansas at Little Rock (UALR) [46] and reported in the manuscript [47].

For the ML models, each iteration involved updated versions of the prototype using different models, testing each of them first with generated data and later with real-world collected data, and each time gathering feedback and revising the artifact accordingly.

Experimental Design typically follows an iterative design process; thus, this work followed different cycles of design. Each cycle is composed of the elements shown in *Figure_1* below:

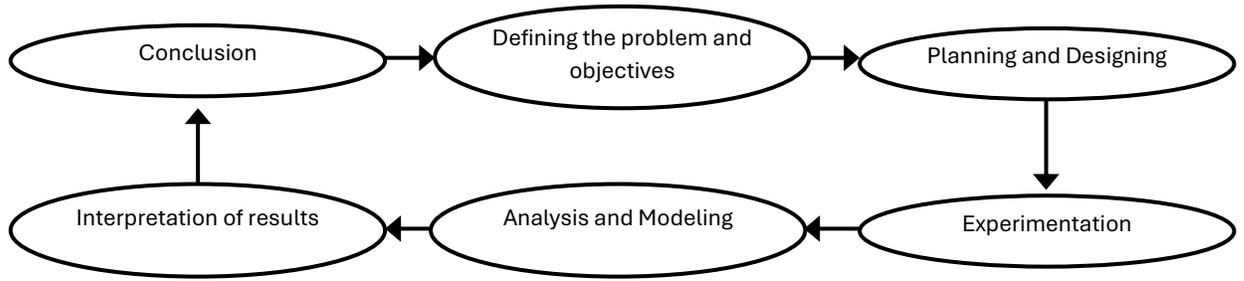

*Figure_1: Experimental Design Research Regulative Cycle*

## 3.2. Experiment

### 3.2.1. General Architecture

The system is composed of three sub-systems as represented in the state machine diagram in *Figure_2*.

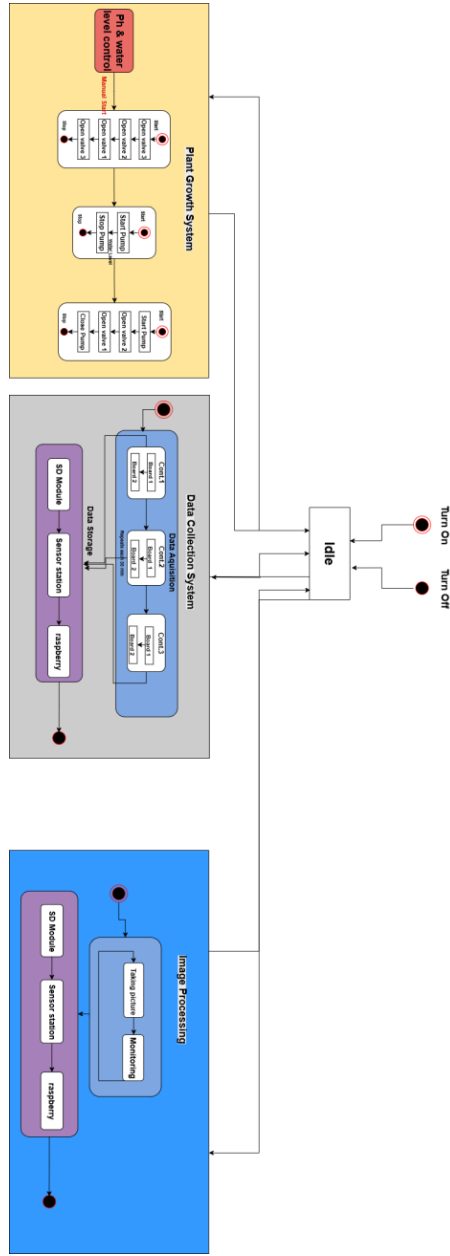

*Figure_2: State Machine Diagram*

The subsystems showcased in the diagram above are described as follows:

**- The Plant Growth System:** This sub-system is designed to measure the climate around the indoor plants, such as the humidity, light intensity, CO2 levels in the air, temperature and the amount of nutrients in the water. Additionally, this sub-system manages the watering system with two different actuators, a relay and a water pump. The water pump is connected to the water source and is activated by the relay when necessary. This sub-system has also a control valve which controls the distribution of water and nutrients as well as another relay for light control.

**- The Image Processing System:** This sub-system is designed to capture pictures of plants, using a camera, and upload them to an image processing model. The aim is to automate the plant growth measurements and make them more precise.

**- The Data Collection System:** This sub-system is designed to predict the growth of plants by collecting and interpreting sensor data. This prediction is based on the comparison of the conditions that our plants are provided with and the conditions that the specific type of plant needs to grow. The prediction is based on three elected Machine Learning models which are Linear Regression, LSTM, and DNN. These models will have time sequence data as input and output predictions. The goal of this prediction is to maximize crop production.

The current prototype of our vertical hydroponic system is shown in *Figure_3* below:

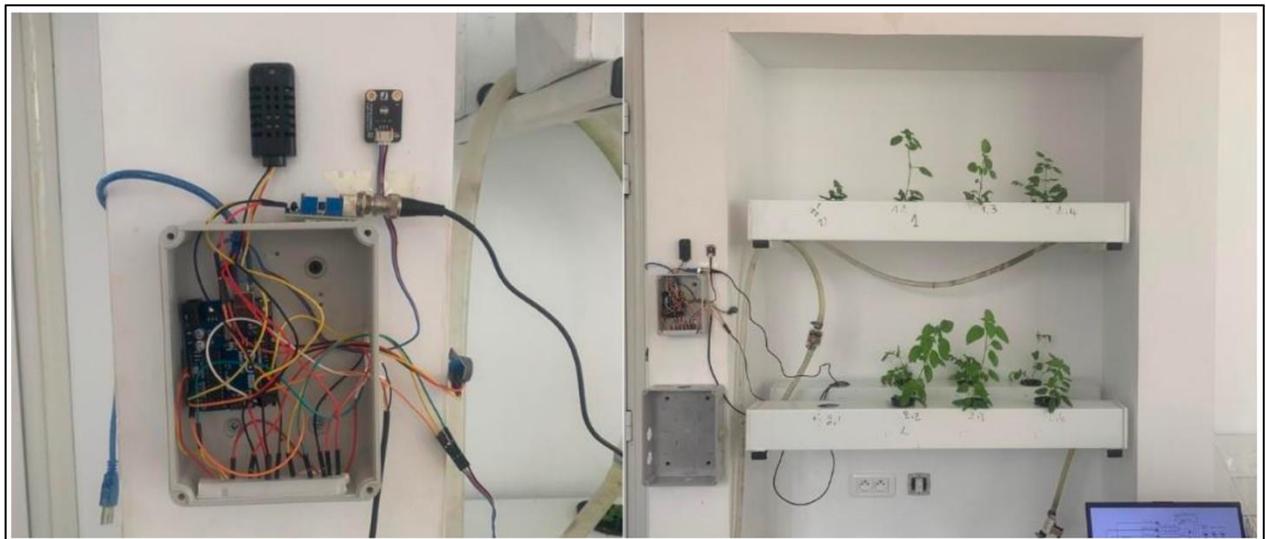

*Figure_ 3: The actual prototype of the vertical hydroponic system*

### 3.2.2. Data collection and control unit

The plant growth and data collection systems combined are considered in this work as the data collection and control unit. The structure of this unit within the vertical hydroponic farm is shown in the *Figure_4* below:

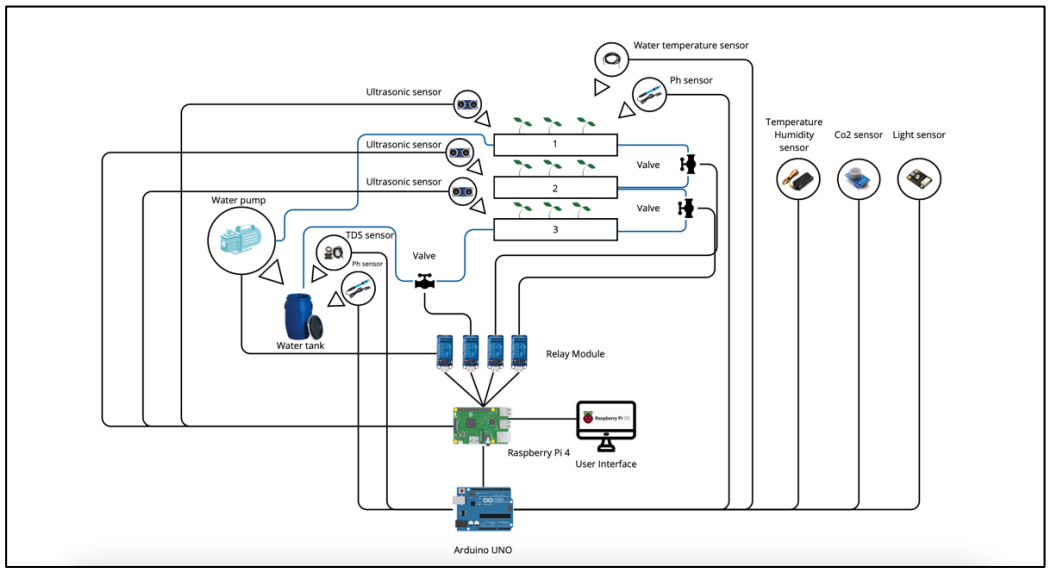

*Figure_4: Architecture of the data collection system*

The system manages water distribution and monitors the environmental parameters for plant growth. At the center of this setup is a Raspberry Pi 4 that is functioning as the central control unit for the water system. Three ultrasonic sensors are connected to the Raspberry Pi and are used to monitor water levels in containers. This architecture was designed to be compatible with low-power microcomputers such as Raspberry Pi 4 for future edge implementation, though current experiments were executed on a simulated environment (Google Colab CPU).

A water pump is linked to the Raspberry Pi via a relay and is situated within the main water tank. When activated, this pump starts the flow of water into the first container, then into the second and third containers. The water flow continues until the desired water level is reached in each container. To regulate the water flow, valves controlled by relays and interfaced with the Raspberry Pi are placed between containers.

Once all the containers reach an optimal water level, the relay is deactivated, and the flow stops. Also, a valve between the third container and the main water tank facilitates water drainage when needed. In addition to the water management system, an Arduino board monitors the environmental parameters critical for plant health. These sensors include a light sensor, a temperature and humidity sensor, a water temperature sensor, a Total Dissolved Solids (TDS) sensor, a pH sensor, and a $CO_2$ sensor. These sensors are positioned throughout the hydroponics' system and collect data.

The Arduino board processes the sensor data and transmits it to the Raspberry Pi, where it is used for system monitoring and control. Finally, a user interface will be linked to the Raspberry Pi and will be used as a simulation platform that displays the findings. This prototype will support a broader understanding of the use of Machine Learning in IoT and its contribution to economic growth in the agriculture sector.

### 3.2.3. Experimental Protocol

Figure 5 shows the crops at the start and end of the experiment.

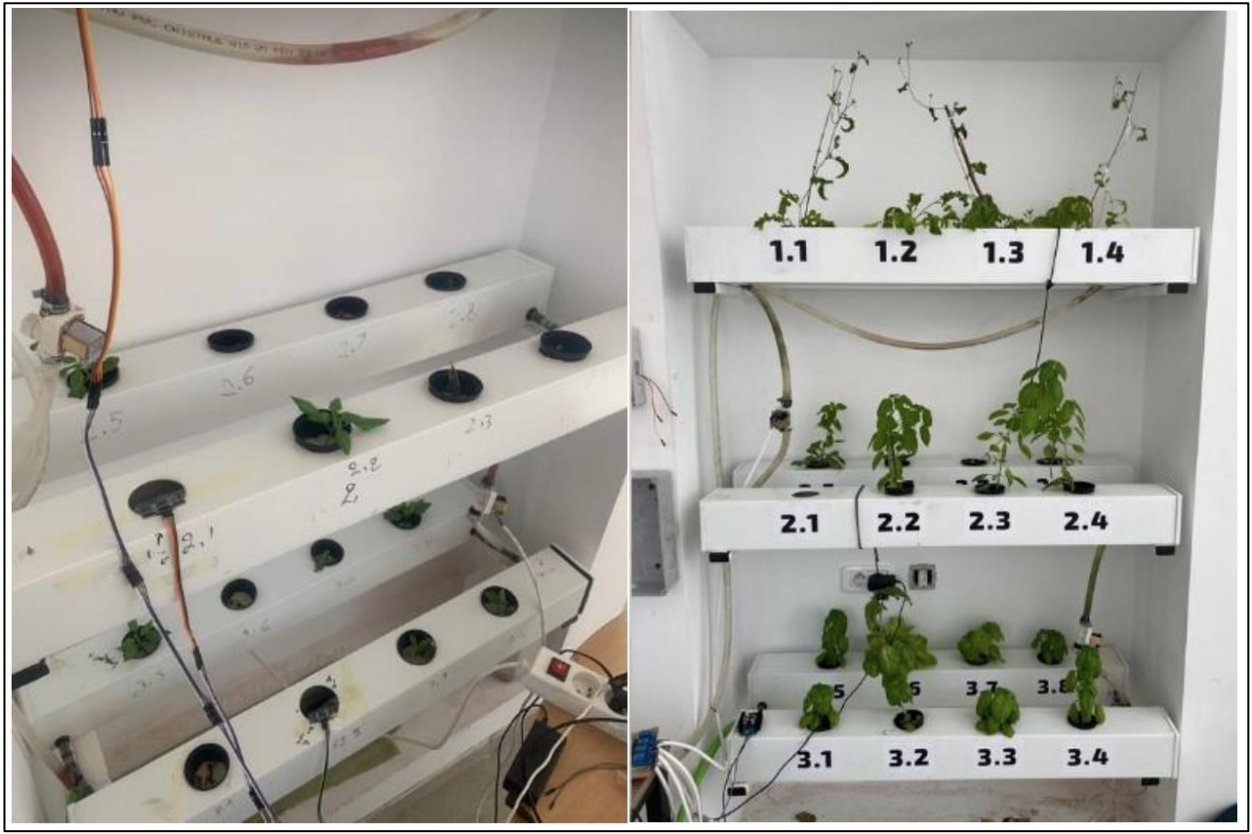

*Figure_5: Planted crops used in the experiment, beginning and end of the experiment*

# 4. Data Collection

## 4.1. Data Approach and Boundaries

Being aware of the constraints related to data collection, we decided on a list of constants and boundaries that aim to guide data analysis and predictive study results. First, for this research, our data is collected only in the spring season and thus, the values are only valid for that season. Second, the data is only collected during daytime, thus, the nighttime part of the plant cycle will not be included.

Third, the fertilizer, labeled for now as "fertilizer A", is a commercial nutrient solution. Thus, the effects of different fertilizers with different compositions will not be part of this research. This study focuses mainly on the effect of fertilizer A on the system as a whole.

So, we recognize the restricted dataset (spring/daytime with one fertilizer and one crop) limits generalizability. Future work will extend experiments to nighttime cycles, multi-season data, and diverse crops/fertilizers, including simulated anomalies to improve robustness.

## 4.2. Data Collection and Preprocessing

Our dataset had around 10k data points from the different sensors of the system. The growth data was collected for 20 consecutive days, with the sensors reading every minute, daily, from around 10AM until 6PM.

This dataset is not publicly available. Researchers who are interested in accessing data may contact the corresponding author to request it.

After the data collection, we explored the dataset and analyzed its values and patterns. A strategy to understand the data is to visualize it and detect patterns or any particularities. The visualization was done by Python code, setting the date and time as the x-axis, the sensor data separately as y-axis, and setting the scale to be specific to each variable.

Details about the used sensors and the graphic representation of their collected data can be found in the appendix from *Figures* 12 to 17.

# 5. Machine Learning models and features

## 5.1. Linear Regression model

### 5.1.1. Linear Regression implementation steps

During the implementation of our Linear Regression model, we started by loading the dataset and preprocessing the data, for example, by replacing commas with dots and removing null values, and converting specific columns to numeric values. Next, we split the data into training and testing sets to ensure that there are enough samples for training. The features were then standardized to center the data around the zero value and scale it to unit variance. Next, we imported the linear model from the sklearn.linear_model library, which we trained until convergence. After that, we made predictions on the test set and visualized its results by comparing the predicted and the actual average values over time using a line plot. Next, the performance of our model was evaluated using Mean Squared Error (MSE), Mean Absolute Error (MAE), and R-squared metrics. We also evaluated the efficiency of the model by measuring its resource usage which are disk usage, execution time, CPU load, and RAM consumption.

### 5.1.2. Parameters of the Linear Regression Model

The Linear Regression model used in this study is characterized by several parameters described in the *Table_2* below:

| | |
|---|---|
| **Feature Scaling** | The features were standardized using the StandardScaler class from scikit-learn. This ensures that all the features have a mean of 0 and a standard deviation of 1, which helps prevent features with larger scales from dominating the model's optimization process. |
| **20-80 Data Splitting** | The dataset was split into training and testing sets using the train_test_split function from scikit-learn. We used the test size of 0.2, which is a common choice in ML that ensures a balance between having enough data for training to learn meaningful patterns and having enough data for testing to accurately evaluate the model's performance. This also helps in preventing any overfitting issues. |

*Table_2: Parameters of the Linear Regression model*

## 7.2. LSTM model

### 7.2.1. LSTM model steps

To implement our LSTM model, we loaded our dataset and preprocessed it. and then split it into both training and test sets. After that, we reshaped the data to fit the LSTM input format, time steps, and features. Next, we defined a sequential Keras model that uses two LSTM layers with dropout regularization and a dense output layer. This LSTM model was trained for 300 epochs and implemented using the Adam optimizer and Mean Squared Error loss function. Similar to all our models, we assessed the performance using MSE, MAE, and R-squared metrics, with the predicted values versus the actual values plotted for visualization. Similarly, the model efficiency was evaluated by the resource usage, including disk, CPU, RAM, and execution time.

### 5.2.2. Parameters of the LSTM model

The LSTM model is characterized by several parameters described in the *Table*_3 below:

| | |
|---|---|
| **Model Architecture** | The architectures were defined using the Sequential class from TensorFlow.keras. The two LSTM layers have 100 units each, followed by dropout layers with a dropout rate of 0.3. The activation function used was ReLU (Rectified Linear Unit). This adds non-linearity to the model and helps in capturing complex temporal patterns in sequential data. |
| **20-80 Data Splitting** | Similarly to our previous model, we used the 80-20 data split. |
| **0.3 Dropout Rate** | We used a dropout rate of 0.3 such that 30% of the neurons in the LSTM layers randomly dropped out during each training epoch. This helped in preventing our model from relying too much on specific temporal features. |

Table_3: Parameters of the LSTM model

## 5.3. DNN model

### 5.3.1. DNN model steps

Like both our other models, our dataset was loaded, preprocessed, split into training and testing sets, and the features were standardized. We created a multi-layer Perceptron regression model with two hidden layers of 100 and 50 neurons created using MLPRegressor and trained incrementally for 50 epochs. Predictions were made on the test set, and results were visualized by plotting predicted versus actual values and the loss curve over iterations. Model performance was evaluated using MSE, MAE, and R-squared metrics. Additionally, disk usage, execution time, CPU load, and RAM consumption were measured to assess resource utilization.

### 5.3.2. Parameters of the DNN Model

The DNN model is characterized by several parameters described in the *Table_4* below:

| | |
|---|---|
| **Model Architecture** | The architecture was defined using the MLPRegressor class from scikit-learn. The activation function that we used in the hidden layers was ReLU (Rectified Linear Unit), which adds non-linearity and helps capture the complex patterns in our data. |
| **20-80 Data Splitting** | The splitting is similar to the one defined for our Linear Regression and LSTM models. |
| **0.3 Dropout Rate** | The dropout is similar to the one defined for our LSTM model. |

*Table_4: Parameters of the LSTM model*

# 6. Performance Assessment and Prediction of Plant Growth Patterns

## 6.1. Evaluation Metrics and Resource Usage

For the evaluation metrics, in general, a lower MSE and MAE indicate better model accuracy while a closer R2 value to 1 indicates a better model fit compared to a value closer to 0.

Moreover, to evaluate the computational resources used by the Linear Regression model, we ran it on the CPU provided by Google Colab, which is the Intel(R) Xeon(R) CPU with vCPUs (virtual CPUs) and 13GB of RAM.

The results of the evaluation metrics and resource usage of each model are summarized in the *Table_5* below:

| Category | Model | Values | Analysis |
|---|---|---|---|
| Evaluation Metrics | Linear Regression | MSE = 0.020 | The squared difference between the predicted and actual values is relatively small. |
| | | MAE = 0.110 | The model's predictions are within approximately 0.11 units of the actual values. |
| | | R2 = 0.185 | The independent variables included in the model contribute 18.54% of the variability observed in the target variable, indicating moderate predictive power. |
| | LSTM | MSE = 0.00002 | The squared difference between the predicted and actual values is very small, indicating high precision in the model's predictions. |
| | | MAE = 0.003 | The average absolute difference between the predicted and actual values is relatively low, indicating a high level of accuracy in the model's predictions. |
| | | R2 = 0.999 | The model explains 99.9% of the variance in the dependent variable, demonstrating a high level of predictive power. |
| | DNN | MSE = 0.00045 | The squared difference between the predicted and actual values is relatively small. This indicates a good level of accuracy in the model's predictions. |
| | | MAE = 0.015 | The model's predictions are 0.015 units away from the actual values. This value indicates that the model's predictions are reasonably close to the true values. |
| | | R2 = 0.981 | The model explains 98.2% of the variance in the dependent variable. This means that the predictions closely follow the variation observed in the actual values. |
| Resource Usage | Linear Regression | Disk Read: 13.85 MB<br><br>Disk Write: 51.59 MB | A moderate disk activity, with a relatively small amount of data being read from disk compared to the amount written. This suggests that the model needed more data storage for its output than for input, which is typical in Machine Learning tasks. |
| | | Execution Time: 11.527 seconds | This measured value is relatively fast, especially knowing that it is executing on a CPU. |
| | | CPU Usage: 85.0% | A significant portion of the CPU's processing power was utilized during model execution. This value seems reasonable for model training. |
| | | RAM Usage: 232.37 MB | The measured value is relatively low. This suggests that the model's memory requirements are moderate. |
| | LSTM | Disk Read: 3.54 MB<br><br>Disk Write: 13.42 MB | A moderate level of disk activity, with a relatively small amount of data being read compared to the amount written. This suggests that, like the Linear Regression, this model requires more storage space for its output than for input. |
| | | Execution Time: 173.650 seconds | This measured value is relatively long but reasonable considering the complexity of the LSTM model and the amount of data being processed. Notably, this execution time is observed while utilizing the CPU for computation. |
| | | CPU Usage: 10.5% | Only a small portion of the CPU's processing power was utilized during model execution. This suggests that there is room for more intensive computations or parallel processing if needed. |
| | | RAM Usage: 822.52 MB | This value is moderately high, indicating that the model's memory requirements are substantial. However, it is within reasonable limits considering the complexity of the LSTM model and the size of the dataset being processed. |
| | DNN | Disk Read: 0.00 MB<br><br>Disk Write: 0.80 MB | A minimal level of disk activity, with a small amount of data being written compared to the amount read. This suggests that the model primarily relies on memory for its operations, which is common in Machine Learning tasks where data is often loaded into memory for processing. |
| | | Execution Time: 33.250 seconds | This measurement is relatively short and reasonable given the complexity of the DNN model and the size of the dataset being processed. |
| | | CPU Usage: 5.0% | Only a small fraction of the CPU's processing power was utilized during model execution. This suggests that the model's computational demands were modest and did not |

|  |  |  | heavily burden the CPU, leaving room for other processes to run concurrently. |
|  |  | RAM Usage: 285.38 MB | This measured value is relatively moderate, indicating that the model's memory requirements are reasonable considering the complexity of the DNN model and the size of the dataset. |

*Table_5: Evaluation Metrics and Resource usage of each ML model*

## 6.2. Features, Parameters and Labels

In a Machine Learning model, the features are the inputs used to make a prediction, and the labels are the inputs we want to predict. For this project, the features are: Light, CO2, TDS, TEMP, HUM, and WaterTemp. The one label, on the other hand, is plant growth, since it is the variable that we want to predict. We are using the average growth of all the planted crops.

The LSTM model is structured with 2 LSTM layers, each containing 100 neurons. To ensure an equitable comparison with the DNN model, that we will discuss later, adjustments were made to align the total number of parameters.

The DNN model was constructed with 3 hidden layers, consisting of 300, 300, and 150 neurons, respectively. We adjusted the model parameters to ensure a comparable number of parameters with the LSTM model for a fair comparison.

The shape of our features and labels, as well as the number of parameters in each model, are listed in the *Table_6* below:

| Category | Model | Values | Analysis |
|---|---|---|---|
| Features, Labels and Parameters | Linear Regression | Features shape: (9948, 6) | This indicates that we have 9948 data points (samples) and 6 features for each data point. So, our feature matrix has 9948 rows and 6 columns. |
| | | Labels shape: (9948, 1) | This indicates that we have 9948 data points (samples) and 1 label for each data point. So, our label vector is a column vector that has 9948 rows. |
| | | Number of parameters: 7 | This shows that we have 6 parameters that correspond to the coefficients for each of the 6 features in our data and 1 parameter that corresponds to the intercept term. |
| | LSTM | Features shape: (9948, 10) | This indicates that we have 9948 data points (samples) and 10 features for each data point. So, our feature matrix has 9948 rows and 6 columns. |
| | | Labels shape: (9948, 1) | This indicates that we have 9948 data points (samples) and 1 label for each data point. So, our label vector is a column vector that has 9948 rows. |
| | | Number of parameters: 124901 | This refers to the total number of parameters in the LSTM model, including weights and biases. |
| | DNN | Total parameters: 137701 | This indicated that the model has a total of 137,701 parameters, which includes the weights and biases of the neural network. |
| | | Features shape: (9948, 6) | This indicates that there are 9,948 data points in the dataset, with 6 features each. |
| | | Labels shape: (9948, 1) | The labels shape indicates that there are 9,948 data points in the dataset, with 1 label each. |

*Table_6: Features, Parameters and Labels of each ML model*

### 6.3. Predicted Vs Actual Average Values

After the implementation of the models, we plotted visual comparisons between the average values predicted by each model and the actual average. Each figure below illustrates the relationship between the actual measured basil growth (from the hydroponic experiment) and the predicted growth values generated by one of the machine-learning models: Linear Regression (Fig 6), LSTM (Fig 7), and DNN (Fig 8).

All three plots share the same axis conventions and visual elements for easy comparison. The X-axis: "Index (test set order)" represents the sequence of data samples in the test set used to evaluate each model. Each index corresponds to one observation of environmental conditions (light, $CO_2$, humidity, temperature, nutrient concentration, etc.) and its associated basil growth measurement. In other words, the x-axis reflects the ordered positions of test samples, not chronological time, though the data were originally collected sequentially during the 20-day experiment. Using this index allows the direct visual comparison of model outputs against actual measurements across all test points.

The Y-axis: "Average Plant Growth (cm)" indicates the average basil growth predicted or measured for each test sample. Growth was calculated as the mean value of the physical plant-height measurements recorded from the multiple crop positions (P1.2 … P3.8) in the hydroponic rack. The unit is centimeters (cm), representing the average height of the basil plants at each sampling point. Typical observed values range between 3.3 cm and 4.2 cm, reflecting the mature growth stage of the plants under indoor hydroponic conditions.

In all graphs, the blue line (Actual) illustrates the measured average plant growth from the IoT-recorded dataset, the orange line (Predicted) represents model-predicted growth values generated from the environmental features. The Light-blue shaded band (95 % interval): statistical 95 % prediction interval, representing the range within which the true value is expected to fall with 95 % probability. It expresses model uncertainty.

Additionally, when the predicted and actual averages are identical, the lines fall on the same spot, indicating a perfect prediction. However, deviations from a common point suggest a difference between the predicted and actual values. Clusters of points away from each other indicate an overestimation or underestimation by the model.

The plot for the Linear Regression model is shown in *Figure_6* below:

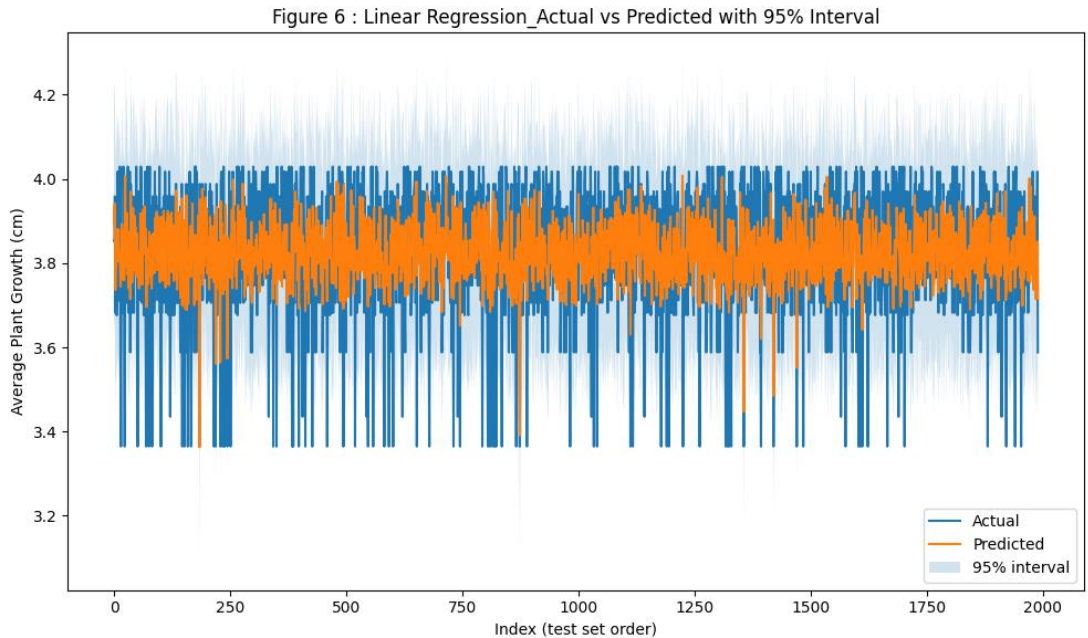

*Figure 6: Actual vs. Predicted basil yield with 95% confidence interval using Linear Regression*

This plot above displays the relationship between the average values predicted by the LR model and the actual average values observed in the dataset. We observe that the predicted and actual values have many common points but also many differences in values. The predicted (orange) curve shows a relatively flat pattern centered around ~3.8 cm, while actual values (blue) fluctuate more widely between 3.3 cm and 4.2 cm. The 95 % interval (shaded band) is quite broad, covering much of the observed variability, an indication of large residual variance and modest predictive power. The model underestimates peaks and fails to capture finer local variations in plant growth.

We can deduce that Linear regression serves as a baseline, it captures the central trend but ignores temporal or interaction effects among predictors. The broad confidence band also illustrates high uncertainty, reinforcing that environmental responses in hydroponics are not linearly separable.

The plot for the LSTM model is shown in the *Figure_7* below:

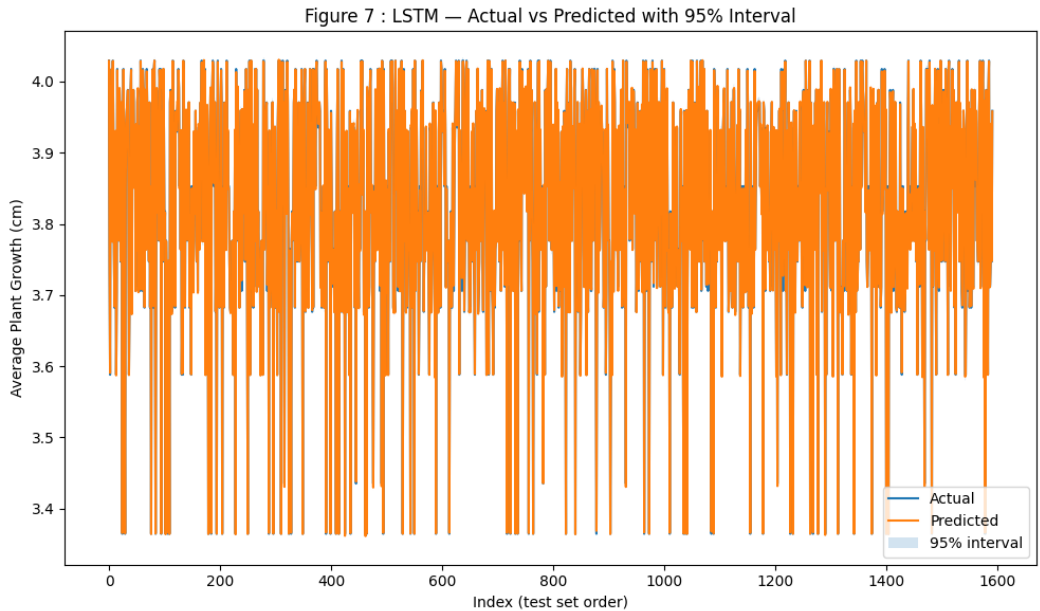

*Figure_7: Actual vs. Predicted basil yield with 95% confidence interval using LSTM*

This plot displays the relationship between the average values predicted by the LSTM model and the actual average values observed in the dataset. Since the previous Linear Regression model used the index as x-axis, for uniformity reasons, we used the same plotting configuration for the LSTM, where the Index of the 'Y_test_original' and 'predictions' arrays was the x-axis of the plot. We see that the predicted and actual values have most of the points in common apart from just a few value differences. The predicted curve (orange) almost entirely overlaps the actual data (blue); both follow the same amplitude and distribution. The confidence interval is narrow and tightly wraps the predicted line, indicating extremely small residual errors and high consistency. Only minor deviations appear at local extremes. This reflects the LSTM's ability to capture temporal dependencies and sequential patterns within sensor readings. The model successfully learns the cyclical trends of environmental parameters and translates them into accurate growth forecasts.

The plot for the DNN model is shown in the *Figure*_8 below:

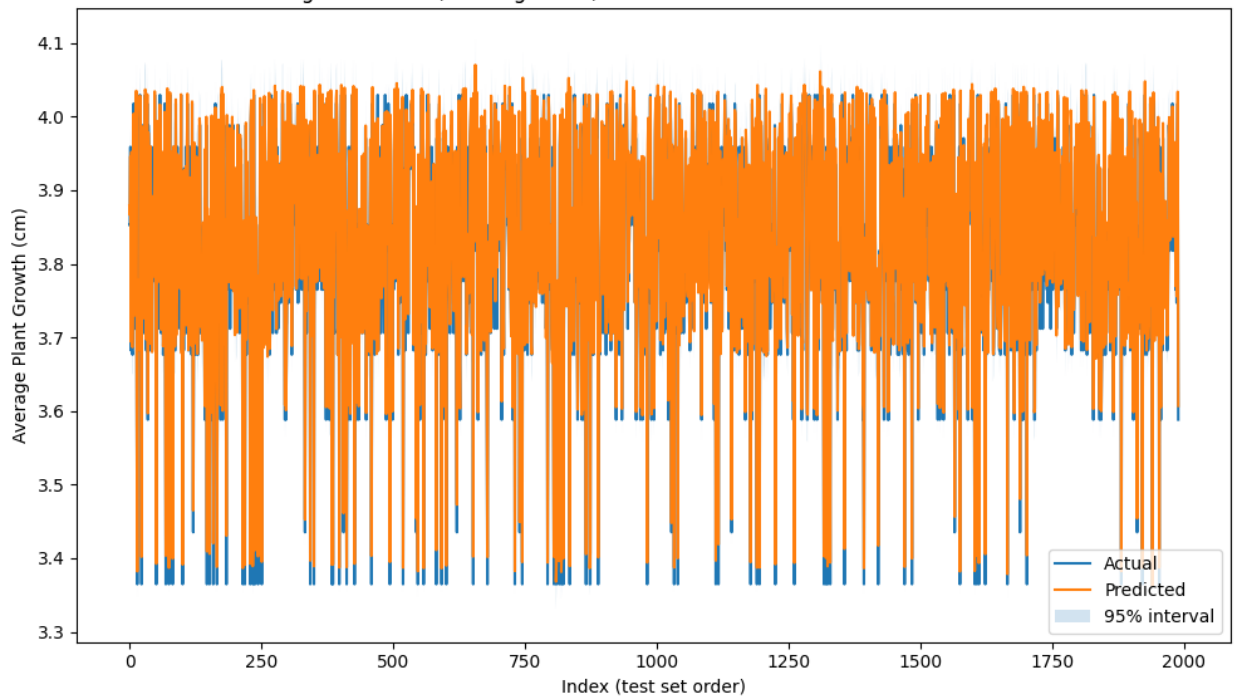

*Figure_8: Actual vs. Predicted basil yield with 95% confidence interval using DNN*

Figure 8 illustrates the correspondence between the average values predicted by the DNN (Deep Neural Network) model and the actual average values observed in the dataset with 95% confidence interval. Overall, we see that most of the predicted and actual values align, with only a few minor differences. The predicted line (orange) aligns closely with actual measurements (blue) but shows slightly more deviation than the LSTM. The confidence band is relatively narrow but wider than for LSTM, suggesting small but non-negligible residuals. And the mean level and amplitude are well preserved.

## 6.4. Feature Sensitivities

To understand how each feature affects each of our models' predictions, we used SHAP (SHapley Additive exPlanations), which is a post hoc explainability method. The following sections show the SHAP analysis results for each model

### 6.4.1. Feature Sensitivity Analysis using SHAP for the Linear Regression Model

We applied SHAP to our trained Linear Regression model which resulted in the plot shown in *Figure_9* below. This graph shows the mean absolute SHAP value for each of the features, which explains its overall contribution to the model's output.

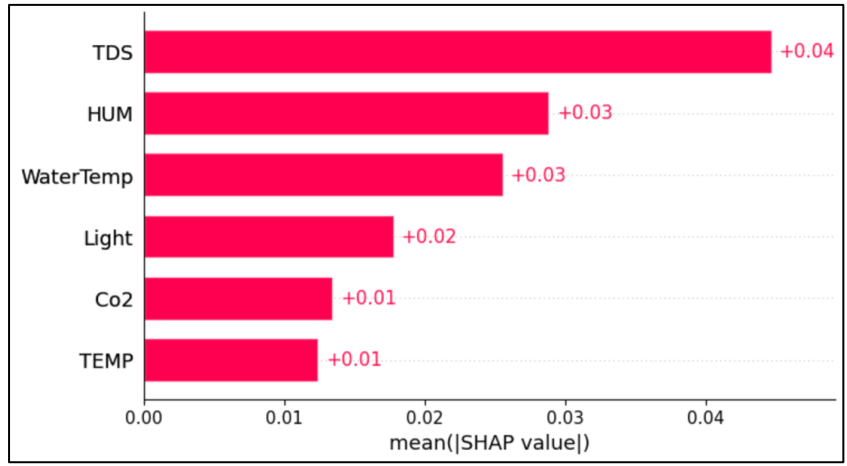

*Figure_9: SHAP Graph applied to the Linear Regression Model*

The analysis of the plot above shows that:

- TDS (Total Dissolved Solids) is the most influential feature. It contributes by 0.04 units to predicted plant growth.
- Humidity (HUM) and Water Temperature (Water Temp) are in second place, each with a mean SHAP value of 0.03.
- Light intensity (Light) has a moderate effect (approximately 0.02), showing its role in photosynthesis. This decreases due to the controlled indoor lighting conditions.
- $CO_2$ concentration (CO2) and ambient temperature (TEMP) show the least impact of approximately 0.01.

### 6.4.2. Feature Sensitivity Analysis using SHAP for the LSTM model

Similar to the Linear Regression feature analysis, and to gain a more thorough understanding of feature importance, we applied SHAP analysis to our LSTM model. The plot in Figure_10 below displays the mean absolute SHAP value for each of our input features and captures its average contribution to the predicted plant growth. Current SHAP analysis provides global feature importance. Future work will integrate temporal SHAP, LIME, and attention mechanisms for instance-level interpretability.

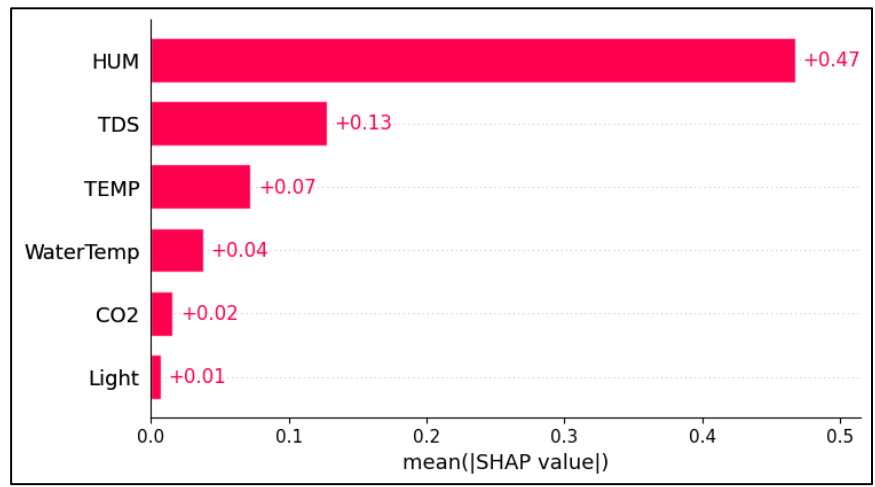

*Figure_10: SHAP Graph applied to the LSTM Model*

This analysis of this plot shows that:

- Humidity (HUM) is the most influential input feature with a SHAP value of 0.31. This shows its importance in predicting plant growth.
- TDS (Total Dissolved Solids) comes in second place and shows moderate contributions with a SHAP value of 0.07.
- Temperature and Water Temperature come in third place, while Light and $CO_2$ show minimal contribution to this output.

### 6.4.3. Feature Sensitivity Analysis using SHAP for the DNN model

Similarly to the previous models, to understand how our input features influence the predictions made by our DNN model, we applied SHAP to our trained model. The plot in *Figure_11* below illustrates the mean absolute SHAP value for each feature, indicating its contribution to our output, which is the predicted plant growth.

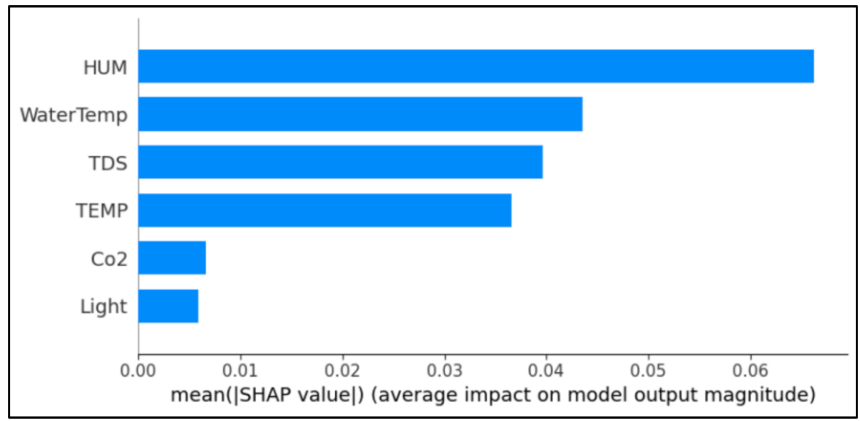

*Figure_11: SHAP Graph applied to the Deep Neural Network (DNN) Model*

This analysis of the plot above shows that:

- Humidity (HUM) is the most influential variable in our DNN model, with a mean SHAP value of more than 0.06.
- Water Temperature (WaterTemp) follows in influence, contributing an important portion to our model's output with an approximate average SHAP value of 0.05.
- TDS (Total Dissolved Solids) comes third in importance, showing its consistent influence on nutrient levels and plant development.
- Ambient Temperature (TEMP) contributes moderately to the model's predictions, likely reflecting environmental control conditions.
- $CO_2$ and Light intensity have the least influence on our predictions, with mean SHAP values less than 0.01, potentially due to the controlled environmental conditions minimizing their variability.

## 7. Results and Comparative Analysis

To get an overall comparison of the three models, we created a table with a comprehensive list of metrics and resource usage across three different models: Linear Regression, LSTM, and DNN. This *Table_7* is shown below:

| Category | Linear Regression | LSTM | DNN |
|---|---|---|---|
| *Parameters* | 7 | 124901 | 137701 |
| *MSE* | 0.020 | 0.00002 | 0.00045 |
| *MAE* | 0.110 | 0.003 | 0.015 |
| *R2* | 0.185 | 0.999 | 0.981 |
| *Execution Time* | 11.527 seconds | 173.650 seconds | 33.250 seconds |
| *CPU Usage* | 85.0% | 10.5% | 5.0% |
| *RAM Usage* | 232.37 MB | 822.52 MB | 285.38 MB |
| *Disk Read* | 13.85 MB | 3.54 MB | 0.00 MB |
| *Disk Write* | 51.59 MB | 13.42 MB | 0.80 MB |

*Table_7: Comparison of the models' metrics and resource usage*

The results obtained from the Linear Regression model (Figure_6) are consistent with the quantitative outcomes presented in Table_7 ( $R^2 \approx 0.18$ ). This confirms that a purely linear relationship is insufficient to capture the complex and nonlinear dependencies between the environmental factors, such as $CO_2$ concentration, light intensity, humidity, and temperature, and the average basil growth represented on the y-axis (in cm). Although the model approximates the overall trend, the wide 95 % prediction interval and the noticeable deviations between the actual and predicted curves highlight its limited explanatory capability.

Conversely, the LSTM model (Figure_7) demonstrates a very narrow confidence band and an almost complete overlap between the actual and predicted values. This indicates a high level of precision and very low variance, which aligns with the reported evaluation metrics ($R^2 \approx 0.999, MAE \approx 0.003$). However, the nearly perfect fit may also suggest a slight tendency toward overfitting, a concern that was addressed by implementing a rigorous 80–20 data-splitting approach to ensure fair model validation.

The LSTM and DNN models have significantly more parameters compared to Linear Regression, indicating greater complexity and capacity for learning complex patterns in the data.

Both LSTM and DNN models achieved substantially lower MSE values compared to Linear Regression, indicating superior predictive accuracy.

Similar to MSE, LSTM and DNN models outperformed Linear Regression in terms of MAE, proving their ability to make more accurate predictions with smaller errors.

The R2 values for LSTM and DNN models are significantly closer to 1 compared to Linear Regression, indicating a better fit to the data and higher predictive power.

LSTM model had the longest execution time, followed by the DNN model, while Linear Regression was the fastest. This suggests that more complex models require more computational time.

Linear Regression used the highest percentage of CPU resources, followed by LSTM and DNN. This reflects the computational intensity of each model, with simpler models requiring more CPU resources.

The LSTM model consumed the highest amount of RAM, followed by DNN and Linear Regression. This indicates the memory requirements of each model, with more complex models requiring more memory.

The amount of data read and written to disk varied across models, with Linear Regression requiring the most disk activity. This suggests different data processing and storage patterns for each model.

Based on these observations, if accuracy and predictive power are the primary concerns, the LSTM model seems to be the best choice. However, if computational efficiency and resource usage are important factors, the DNN model would be preferred due to its lower CPU and RAM usage while still providing reasonable accuracy. The Linear Regression model would be suitable for tasks where simplicity and speed are prioritized over accuracy.

The recorded execution times and resource usage correspond to simulation on a general-purpose CPU environment, not on embedded hardware. These benchmarks guide our selection of models for subsequent deployment choice:
- LR: quick estimation on microcontrollers or dashboards.
- DNN: balanced model for continuous monitoring on edge devices.
- LSTM: best for centralized predictive analytics requiring time-series fidelity.

# 8. Conclusions and Future Directions

## 8.1. Addressing the Main Research Question

This study demonstrates that Machine Learning models, when properly integrated with IoT-based environmental monitoring, can accurately forecast basil crop growth in a vertical hydroponic context. The experimental setup collected continuous multi-sensor data: including light intensity, $CO_2$ concentration, ambient temperature, relative humidity, nutrient concentration (TDS), and water temperature, forming a comprehensive dataset that reflects the dynamic plant–environment interaction.

By preprocessing, standardizing, and training ML algorithms on this IoT data, the system successfully generated predictive models capable of estimating basil growth (in cm) with varying degrees of precision, as visualized in the results figures. The findings confirm that ML algorithms can serve as effective predictive tools in precision agriculture, enabling automated decision support and real-time optimization of growing conditions.

**Sub-question 1: Critical Data Points for Prediction**

The analysis identified six key environmental parameters as critical predictors for basil growth: light intensity, $CO_2$ concentration, temperature, humidity, TDS, and water temperature. These features directly influence photosynthesis, nutrient absorption, and transpiration, processes that drive plant development in controlled environments.

Feature sensitivity analyses using SHAP further supported the selection, showing that humidity, TDS, and water temperature had the highest influence across models, with light and $CO_2$ contributing moderately due to the system's controlled indoor lighting and $CO_2$ regulation.

The inclusion of these parameters ensured that the models captured both environmental variability and physiological responses, which are essential for reliable growth prediction.

**Sub-question 2: Preparing IoT Data for Machine Learning**

The IoT data collected from the hydroponic setup required systematic data preprocessing before being used in ML modeling. The steps included:

- **Data cleaning:** handling missing and inconsistent readings, converting commas to decimal points, and ensuring numerical consistency across sensor outputs.
- **Feature scaling**: applying standardization to center features around zero mean and unit variance, which is critical for gradient-based models such as LSTM and DNN.
- **Data splitting:** dividing the dataset using an 80–20 training–testing ratio to prevent bias and evaluate generalization performance.
- **Feature engineering:** for time-series models (LSTM), temporal attributes such as Year, Month, Day, and Hour were extracted from timestamps to enable sequential pattern recognition.

This pipeline transformed raw, heterogeneous IoT data into structured numerical inputs suitable for ML, ensuring both interpretability and computational efficiency. The approach confirms that data preprocessing is foundational to integrating IoT systems with AI-based predictive analytics in agriculture.

**Sub-question 3: Comparative Performance of Linear Regression, LSTM, and DNN**

A comparative analysis of the three predictive models highlights the trade-offs between interpretability, computational cost, and predictive accuracy:

Linear Regression (Figure 6) achieved an ($R^2 \approx 0.18$), indicating limited capacity to capture the nonlinear and multivariate dependencies between environmental variables and basil growth. While its computational efficiency and interpretability make it suitable for simple applications or quick estimation tasks, its predictive reliability remains low due to high variance and broad confidence intervals.

LSTM (Figure 7) provided the most accurate results ($R^2 \approx 0.999, MAE \approx 0.003$) with nearly perfect alignment between predicted and actual values and an extremely narrow 95 % confidence band. Its sequential architecture effectively modeled temporal dependencies in the IoT data, capturing patterns related to diurnal and environmental cycles. The model's high accuracy confirms the importance of temporal learning in hydroponic growth forecasting, though its complexity and higher computational demands may limit real-time deployment on constrained devices.

DNN (Figure 8) achieved a strong predictive performance ($R^2 \approx 0.981$, $MAE \approx 0.015$) while maintaining lower computational requirements compared to LSTM. It effectively learned nonlinear interactions between environmental parameters without explicitly considering temporal sequences. This balance between accuracy and efficiency makes DNNs highly suitable for real-time, resource-aware edge computing environments.

The performance hierarchy, LSTM>DNN≫Linear Regression, confirms that deep learning models outperform traditional regression techniques for IoT-based crop prediction due to their ability to capture complex and temporal relationships.

The findings collectively show that Machine Learning can transform IoT-enabled hydroponic systems into intelligent predictive platforms. By continuously learning from environmental data, such models can anticipate growth trends, optimize nutrient dosing, and adjust lighting or irrigation schedules automatically. The combination of IoT sensing and deep learning provides a scalable foundation for data-driven precision agriculture, supporting sustainable production and resource efficiency.

### 8.2. Synthesis and Future Directions

This study explores the use of ML models with IoT data from a vertical hydroponic farm to predict basil growth. It's based on the comparison of three Machine Learning models that are Linear Regression, LSTM and DNN using data collected from an indoor vertical farming unit located at the Mediterranean Institute of Technology (MedTech) in Tunisia. Although the project has several groups working on different parts of it, this research is about the Machine Learning part of it conducted at the University of Arkansas at Little Rock (UALR).

After analyzing each ML model, we found that the LSTM model has promising accuracy and predictive power. This model was able to capture the complex time dependencies and patterns present in our data and generally performs well with sequential data such as time series forecasting. Even though this model had a longer execution time and higher resource usage compared to the DNN model, its high predictive capabilities justify its use in scenarios where the predictions must be accurate, and the computational resources are not limited.

Moreover, the performance of the DNN model showed a balance of accuracy and resource usage efficiency. DNNs are flexible and scalable which explains the high accuracy results while consuming less computational resources than the LSTM model. These results make this model well-suited for applications favoring computational efficiency over prediction power, such as large-scale data processing systems.

Overall, the choice between the three models, Linear Regression, LSTM and DNN, depends on the specific constraints and requirements of the application. While the simplicity and high speed of the Linear Regression model is suitable for rapid processing and applications requiring basic predictive tasks, the LSTM model provides high prediction powers at the cost of high resource usage. Alternatively, the DNN model is a good alternative for scenarios where accuracy and efficiency are equally important. The addition of 95 % bands enhances transparency by showing model uncertainty, essential for real-world decision support in automated farming. Together, they demonstrate a clear progression from basic statistical regression to advanced deep learning, illustrating how temporal and nonlinear modeling dramatically improves the precision of basil yield prediction in IoT-enabled hydroponic systems.

Future work will focus on enhancing model generalization through cross-validation, early stopping, and L2 regularization. Beyond Linear Regression, LSTM, and DNN, additional benchmarking will consider Random Forest, XGBoost, ARIMA, and CNN models. Planned extensions also include integrating an image-processing module for automated plant growth assessment, incorporating a new fertilizer formulation developed by the MedTech Chemistry Department, and enriching the dataset with nighttime and multi-season measurements, as well as additional plant species. These improvements aim to overcome the current dataset's limitations, restricted to spring, daytime, and a single fertilizer, by introducing greater diversity and simulated anomalies to improve robustness. Future iterations will also integrate image-based features, nutrient uptake rates, and time-lagged variables into the prediction pipeline. Finally, forthcoming work will explore deployment of the trained models on embedded platforms such as Raspberry Pi 4 to evaluate real-time inference latency, model compression, and energy efficiency. All current experiments were conducted in a simulated CPU environment; hardware-based evaluation will represent the next phase of this research.

## 9. Data availability

The dataset and code are now publicly available at:
https://github.com/emnabouzid13/Machine_Learning_Based_Basil_Yield_Prediction_in-IoT_Enabled_Indoor_Vertical_Hydroponic_Farms

## 10. Conflicts of interest

The authors declare that there is no conflict of interest regarding the publication of this article.

## 11. Co-Author contributions: CRediT

- Baccar Noura: Project administration, Supervision, Writing – review & editing
- Iqbal Kamran: Supervision, Conceptualization, Writing – review & editing
- Chaouch Yassine: Validation, Resources, Writing – review & editing
- Ben Youssef Fares: Hardware, Software, Data curation
- Regayeg Amine : Hardware, Software, Data curation
- Toumi Sarra: Resources, Investigation
- Nsir Houda : Conceptualization, Resources
- Mseddi Amina: Investigation, Conceptualization

- Costelle Leila, Conceptualization, Funding acquisition

## 12. Funding sources

This study was funded by the Mediterranean Institute of Technology (MedTech) and the Research Center at South Mediterranean University (SMU) as part of their strategic vision to support the Sustainable Development Goals, particularly Goals 2 (Zero Hunger) and 11 (Sustainable Cities and Communities). Their contributions promote innovative agricultural technology research.

## 13. Acknowledgment

This paper is based on work originally presented in the master thesis titled *Design and Implementation of a Machine Learning Predictive Model for Crop Yield in Indoor Vertical Hydroponics Farming*. We would like to acknowledge the insights and foundational research from this thesis that have contributed significantly to the development and findings of this paper.

The authors acknowledge the use of generative AI tools (ChatGPT by OpenAI) in the drafting and refinement of this manuscript. The AI assistance was limited to language editing and improving clarity. All content was reviewed and verified by the authors, who take full responsibility for the final version of the manuscript.

## 14. Glossary

**Vertical farming:** Vertical Farming is a cutting-edge agricultural approach that revolutionizes traditional cultivation methods where crops are planted in layers that are stacked vertically, usually in indoor environments. The primary advantage of vertical farming is its efficient use of space, which allows for increased crop yields in smaller spaces, making it valuable in urban areas where land is limited.

**Hydroponics:** Hydroponics are based on cultivating plants without soil. This technology uses the fact that plants can still thrive when supplied with all essential nutrients in a non-organic form, through a liquid solution, without a solid medium. Moreover, hydroponic systems are well suited for high value vegetables because they are a way to avoid soil-born pests. Hydroponic farming is also an efficient way of growing plants indoors and helping in the production of nutrient-rich plants faster compared to traditional farming.

**Internet of Things (IoT):** In the context of hydroponics and indoor farming, the Internet of Things (IoT) adds a revolutionary technological environment. Through a network of sensors, actuators, and control systems, IoT optimizes the cultivation process. Such systems enable precise monitoring and management of environmental variables like nutrient levels, temperature, humidity, and lighting conditions. In the context of hydroponics, IoT enhances efficiency and contributes to sustainable farming by minimizing resource waste and maximization of yield in controlled indoor environments. This combination of advanced technology and agriculture creates an example of the potential for IoT in addressing the challenges of modern food production.

**Machine Learning and Regression:** In the context of hydroponics and indoor farming, ML models can be used to analyze big datasets collected by sensors as well as detect implicit patterns and correlations and provide insights into optimal conditions for plant growth.

ML creates adaptive and dynamic systems in hydroponics, where it learns from real-time data to adjust parameters like environmental factors or water, or nutrient levels. The predictive capabilities of ML also contribute to dynamic decision-making, predicting possible, and mitigating problems. This project involves regression rather than classification in ML, since the focus is on predicting continuous variables related to the hydroponic system, which is plant growth rate. Regression models are well-suited for forecasting and optimizing numerical outputs, which aligns with the goal of using hydroponic conditions for maximum crop productivity.

**Linear Regression:** Linear Regression is a Machine Learning model suitable for modeling relationships between variables that are characterized by continuous numerical values. The key components of a Linear Regression model are:
- The input features which are the variables used for prediction in the regression analysis.
- The regression coefficients which are the parameters that the model learns to assign to
each input feature, indicating their weight in the prediction process.
- The output layer which generates the final prediction based on the input features and their corresponding coefficients.

The primary formula for prediction using Linear Regression involves a linear relation between the input features and the regression coefficients, followed by an optional addition of a bias term.

**Long short-term memory (LSTM):** LSTM is a Machine Learning model that is suitable for handling long-term dependencies in sequential data like speech, time series, and natural language data. The key components of an LSTM network are:
- The input layer which receives input data at each time step.
- LSTM cells are the main building blocks of the network. They are used to select updates, reset, and output information each time.
- The output layer which generates the final output of the network by receiving the final internal state.

Each LSTM cell has these main components:
- The forget gate which decides and selects the information to be removed from the cell at each time step.
- The input gate which controls and selects what new information to be added to the cell at each time step.
- The output gate which passes the updated information from the current time step to the next time step.

**Deep Neural Networks (DNN):** DNN is a Machine Learning model that is suitable for extracting features and patterns from complex datasets, which makes them ideal for tasks involving computer vision, speech recognition, etc. With the current growth of big data, DNNs have become crucial tools for analyzing vast amounts of information and deriving insights. The key components of a DNN network are:

- The input layer serves as the first point of interaction between the network and the external data. It consists of neurons each corresponding to a feature of the input data. The values of these neurons are the input data themselves. The size of this layer is determined by the dimension of the input data.
- The hidden layers which are the intermediate layers between the input and the output layers and are where the actual computation happens. These hidden layers extract features from the input data.
- The output layer which produces the final output of the network. This layer consists of one or more neurons, depending on the nature of the problem being addressed, for example classification or regression. In regression, each neuron in the output layer represents a numerical value. The activation function applied to the neurons in the output layer depends on the task.

**Performance Metrics:** Different evaluation metrics for regressive ML models exist, examples are: Mean Absolute Error (MAE), Mean Squared Error (MSE), and R-squared (R2). These metrics are used to evaluate the performance of a model by calculating the difference between the predicted and the actual values. Each of these metrics serves as a tool to measure the efficacy of models and make decisions about model selection and refinement.

# References

[1] USAID, "Climate Risk Profile Tunisia," Oct. 2018. [Online]. Available: https://www.climatelinks.org/sites/default/files/asset/document/Tunisia_CRP.pdf. [Accessed: Jul. 3, 2025].

[2] P. Arrojo-Agudo, "End of Mission Statement by the Special Rapporteur on the human rights to safe drinking water and sanitation," United Nations Human Rights Special Procedures, 28 Jul. 2022.

[3] H. Knaepen, *Climate risks in Tunisia: Challenges to adaptation in the agri-food system*, European Centre for Development Policy Management (ECDPM), Feb. 2021. [Online]. Available: https://www.cascades.eu/wp-content/uploads/2021/02/Climate-risks-in-Tunisia-Challenges-to-adaptation-in-the-agri-food-system-1.pdf. [Accessed: Jul. 3, 2025].

[4] A. Elmahdi, A. Badawy, and H. A. P. Lopez, *Addressing the water challenges in the agriculture sector in Near East and North Africa*, Food and Agriculture Organization of the United Nations, 2022, p. 88. [Online]. Available: https://openknowledge.fao.org/server/api/core/bitstreams/d21c6fbe-8f96-4f7b-b078-55de485ef7c5/content. [Accessed: Jul. 3, 2025].

[5] C. Laura, P. Alessandro, Z. Ilaria, M. Davide, M. Michael, P. Giuseppina, G. Giorgio and O. Francesco, "Improving water use efficiency in vertical farming: Effects of growing systems, far-red radiation and planting density on lettuce cultivation," *Agricultural Water Management,* vol. 285, 2023.

[6] Z. Zhang, R. Michel, and F. Hosseinian, "A comprehensive review on sustainable industrial vertical farming using film farming technology," *Sustainable Agriculture Research*, vol. 10, pp. 46–46, 2020, doi: 10.5539/sar.v10n1p46.

[7] F. Modu, A. Adam, F. Aliyu, A. Mabu, and M. Musa, "A survey of smart hydroponic systems," Advances in Science, Technology and Engineering Systems Journal, vol. 5, no. 1, pp. 233–248, 2020.

[8] A. Dudwadkar, T. Das, S. Suryawanshi, T. Kothawade, and R. Dolas, "Automated hydroponics with remote monitoring and control using IoT," *International Journal of Engineering Research & Technology (IJERT)*, vol. 9, no. 6, Jun. 2020.

[9] H. M. Shetty, K. P. Kshama, N. Mallya, and P. [Last name if available], "Fully automated hydroponics system for smart farming," *International Journal of Engineering and Manufacturing (IJEM)*, vol. 11, no. 4, pp. 33–41, 2021, doi: 10.5815/ijem.2021.04.04.

[10] A. Mehboob, W. Ali, T. Rafaqat, and A. Talib, "Automation and control system of EC and pH for indoor hydroponics system," in *Proc. 4th Int. Electrical Engineering Conf. (IEEC)*, Karachi, Pakistan, Dec. 2019, pp. 1–6. [Online]. Available: https://ieec.neduet.edu.pk/2019/Papers_IEEC_2019/IEEC_2019_33.pdf. [Accessed: Jul. 3, 2025].

[11] Mduma, N., Mayo, F., 2024. Updating machine learning imagery dataset for maize crop: A case of Tanzania with expanded data to cover the new farming season. Data in Brief, vol. 54. https://doi.org/10.1016/j.dib.2024.1103599

[12] Idoje, G., Mouroutoglou, C., Dagiuklas, T., Kotsiras, A., Muddesar, I., & Alefragkis, P. (2023). Comparative analysis of data using machine learning algorithms: A hydroponics system use case. *Smart Agricultural Technology, 4*, 100207. https://doi.org/10.1016/j.atech.2023.100207

[13] N. Jaybhaye, P. Tatiya, A. Joshi, S. Kothari and J. Tapkir, "Farming Guru: - Machine Learning Based Innovation for Smart Farming," 2022 4th International Conference on Smart Systems and Inventive Technology (ICSSIT), Tirunelveli, India, 2022, pp. 848-851, doi: 10.1109/ICSSIT53264.2022.9716287.

[14] T. M. Bandara, W. Mudiyanselage and M. Raza, "Smart farm and monitoring system for measuring the Environmental condition using wireless sensor network - IOT Technology in farming," 2020 5th International Conference on Innovative Technologies in Intelligent Systems and Industrial Applications (CITISIA), Sydney, Australia, 2020, pp. 1-7, doi: 10.1109/CITISIA50690.2020.9371830.

[15 ] J. Choi, D. Lim, S. Choi, J. Kim and J. Kim, "Light Control Smart Farm Monitoring System with Reflector Control," 2020 20th International Conference on Control, Automation and Systems (ICCAS), Busan, Korea (South), 2020, pp. 69-74, doi: 10.23919/ICCAS50221.2020.9268238.

[16] A. Calcante, R. Oberti, and F. M. Tangorra, "Definition of linear regression models to calculate the technical parameters of Italian agricultural tractors," Journal of Agricultural Engineering, 2023, doi: 10.4081/jae.2023.1525.

[17] A. Jain, "Ridge and Lasso Regression in Python | Complete Tutorial (Updated 2024)," 11 Jan 2024. [Online]. Available: https://www.analyticsvidhya.com/blog/2016/01/ridge-lassoregression-python-complete-tutorial/.

[18] M. Clyde, "An introduction to Bayesian thinking," Jun. 15, 2022. [Online]. Available: https://statswithr.github.io/book/introduction-to-bayesianregression.html

[19] Leon, M., Cornejo, G., Calderón, M., González-Carrión, E., & Florez, H. (2022). Effect of Deforestation on Climate Change: A Co-Integration and Causality Approach with Time Series. Sustainability, 14(18), 11303. https://doi.org/10.3390/su141811303

[20] D. M. S. W. Agni Isador Harsapranata, "Liner Regression Analysis In Forecasting Hydroponic Plant Nutrient Needs," JURNAL INFOKUM, vol. Volume 10, no. No 5, 2022.

[21] H. Helmy, D. A. M. Janah, A. Nursyahid, M. N. Mara and T. A. Setyawan, "Nutrient Solution Acidity Control System on NFT-Based Hydroponic Plants Using Multiple Linear Regression

[22] T. van Klompenburg, A. Kassahun, and C. Catal, "Crop yield prediction using machine learning: A systematic literature review," *Computers and Electronics in Agriculture*, vol. 177, p. 105709, 2020, doi: 10.1016/j.compag.2020.105709.

[23] J. Kim, B. Yu, and S. O'Hara, "LSTM filter for smart agriculture," *Procedia Computer Science*, vol. 210, pp. 289–294, 2022, doi: 10.1016/j.procs.2022.10.152.

[24] Suebsombut, P., Sekhari, A., Sureephong, P., Belhi, A., & Bouras, A. (2021). Field Data Forecasting Using LSTM and Bi-LSTM Approaches. *Applied Sciences*, *11*(24), 11820. https://doi.org/10.3390/app112411820

[25] B. Yu and J. Kim, "Using a neural network to detect anomalies given an N-gram profile," *arXiv preprint* arXiv:2104.05571, 2021. doi: 10.48550/arXiv.2104.05571.

[26] X.-Y. Liu, Y. Fang, L. Yang, Z. Li, and A. Walid, "Chapter 9 - High-performance tensor decompositions for compressing and accelerating deep neural networks," in *Tensors for Data Processing*, Y. Liu, Ed. Academic Press, 2022, pp. 293–340, doi: 10.1016/B978-0-12-824447-0.00015-7.

[27] Y. LeCun, Y. Bengio, and G. Hinton, "Deep learning," *Nature*, vol. 521, no. 7553, pp. 436–444, May 2015, doi: 10.1038/nature14539.

[28] T. van Klompenburg, A. Kassahun, and C. Catal, "Crop yield prediction using machine learning: A systematic literature review," Computers and Electronics in Agriculture, vol. 177, p. 105709, 2020, doi: 10.1016/j.compag.2020.105709.

[29] B. P. S. G. a. S. C. Abhasha Joshi, "Remote-Sensing Data and Deep-Learning Techniques in Crop Mapping and Yield Prediction: A Systematic Review," Remote Sensing, vol. 15(8), 2023.

[30] W. Samek, T. Wiegand, and K.-R. Müller, "Explainable artificial intelligence: Understanding, visualizing and interpreting deep learning models," *ITU Journal: ICT Discoveries*, no. Special Issue No. 1, 2017.

[31] M. A. Iqbal, "Application of Regression Techniques with their Advantages and Disadvantages," Power Systems Simulation Research Lab, 2021.

[32] J. Abate, A. Aman, and D. Adem, "Integration of satellite data for predicting crop yields in Eastern Ethiopia using machine learning," Scientific Reports, vol. 15, Art. 33809, 2025. DOI: 10.1038/s41598-025-00810-z.

[33] Z. A. Salami, "Modelling crop yield prediction with Random Forest and remote sensing data," Natural and Engineering Sciences, vol. 68, 2025. doi:10.28978/nesciences.1763843.

[34] A. Chlingaryan, S. Sukkarieh, and B. Whelan, "Machine learning approaches for crop yield prediction and nitrogen status estimation in precision agriculture: A review," Computers and Electronics in Agriculture, vol. 151, pp. 61–69, 2018.

[35] P. Zhang, L.-Y. Li, J. Wang, et al., "Ensemble Learning for Oat Yield Prediction Using Multi-Stage UAV Multispectral Data," Remote Sensing, vol. 16, no. 23, Art. 4575, 2024. DOI: 10.3390/rs16234575.

[36] B. Mafrebo Lionel, R. Musabe, O. Gatera, and C. Twizere, "A comparative study of machine learning models in predicting crop yield," Machine Learning and Applications, vol. 3, no. 2, pp. 145–158, 2025. DOI: 10.1007/s44279-025-00335-z.

[37] G. Yu, M. Wu, and J. Chen, "Global de-trending significantly improves the accuracy of XGBoost in yield forecasting," Journal of Applied Remote Sensing, vol. 18, no. 1, pp. 210–222, 2024.

[38] Y. Tripathi, P. Bhatnagar, and R. N. Jain, "Time-Series Analysis of Crop Yield Using ARIMA and LSTM Models Under Climate-Change Conditions," Environmental Modelling & Software, vol. 165, p. 105732, 2023

[39] Petteri Nevavuori, Nathaniel Narra, and Tarmo Lipping. 2019. Crop yield prediction with deep convolutional neural networks. Comput. Electron. Agric. 163, C (Aug 2019). https://doi.org/10.1016/j.compag.2019.104859

[40] T K Nida Fariz, S Sharief Basha, "Enhanced crop yield prediction using a hybrid artificial neural network optimized by the Coati optimization algorithm," Results in Engineering, vol. 28, 2025, 107529, ISSN 2590-1230, https://doi.org/10.1016/j.rineng.2025.107529

[41] Jovanovic, Luka & Zivkovic, Miodrag & Bacanin, Nebojsa & Dobrojević, Miloš & Simic, Vladimir & Sadasivuni, Kishor & Tirkolaee, Erfan. (2024). Evaluating the performance of metaheuristic-tuned weight agnostic neural networks for crop yield prediction. Neural Computing and Applications. 36. 10.1007/s00521-024-09850-4.

[42] Alharbi, Amal & H.Rizk, Faris & Sh.Gaber, Khaled & Eid, Marwa & El-kenawy, El-Sayed & Khodadadi, Ehsan & Khodadadi, Nima. (2025). Hybrid deep learning optimization for smart agriculture: Dipper throated optimization and polar rose search applied to water quality prediction. PLOS One. 20. 10.1371/journal.pone.0327230.

[43] Banerjee, S., Nandi, T., Sati, V. P., Mezlini, W., Alkhuraiji, W. S., Al-Halbouni, D., & Zhran, M. (2025). Integrating Remote Sensing, Landscape Metrics, and Random Forest Algorithm to Analyze Crop Patterns, Factors, Diversity, and Fragmentation in a Kharif Agricultural Landscape. *Land*, *14*(6), 1203. https://doi.org/10.3390/land14061203

[44] B. Jones, "What is experimental design?," [Online]. Available: https://www.jmp.com/en_is/articles/what-is-experimental-design.html. [Accessed: Jul. 3, 2025].

[45] Mediterranean Institute of Technology (MedTech), "Homepage," [Online]. Available: https://www.smu.tn/medtech. [Accessed: Jul. 3, 2025].

[46] University of Arkansas at Little Rock, (UALR), "Homepage," [Online]. Available: https://uasys.edu/campuses-units/universities/little-rock/. [Accessed: Jul. 3, 2025].

[47] Bouzid, E. (2024). Design and implementation of a machine learning predictive model for crop yield in indoor vertical hydroponics farming (Order No. 31243476). (3054651537). Retrieved from https://www.proquest.com/dissertations-theses/design-implementation-machine-learning-predictive/docview/3054651537/se-2

# Appendix:

The data received from the sensors is described below:

- **Light:** The light sensor was used to measure the intensity of the light surrounding the plants. Low light levels typically vary between 200-500 lux. Medium light levels range between 500 and 2000 lux, which provides good illumination for photosynthesis and growth. High light levels, surpassing 2000 lux, may be too much light which could potentially lead to stress or damage to the plants. The data received from the light sensor is visualized in the *Figure_12* below:

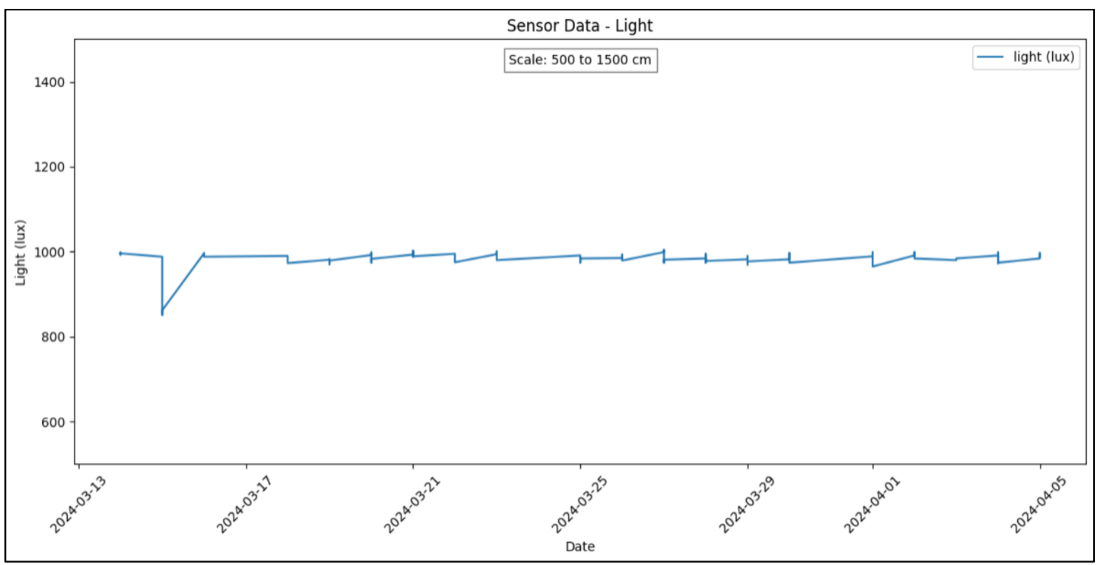

*Figure_12: Data collected from the light sensor*

- **CO2 (Carbon Dioxide)**: The CO2 sensor was used to detect the amount of CO2 surrounding plants. Low CO2 levels are typically considered to be below 400ppm, medium CO2 levels are between 400 and 1,000 ppm. In some cases, this range is considered acceptable for indoor environments. High CO2 levels are usually above 1,000 ppm. Such concentrations are usually caused by poor ventilation or other environmental issues. For controlled-environment agriculture, higher levels of CO2 (up to 1,500-2,000 ppm) are usually preferred to enhance plant growth. The data received from the CO2 sensor is visualized in the F*igure_13* below:

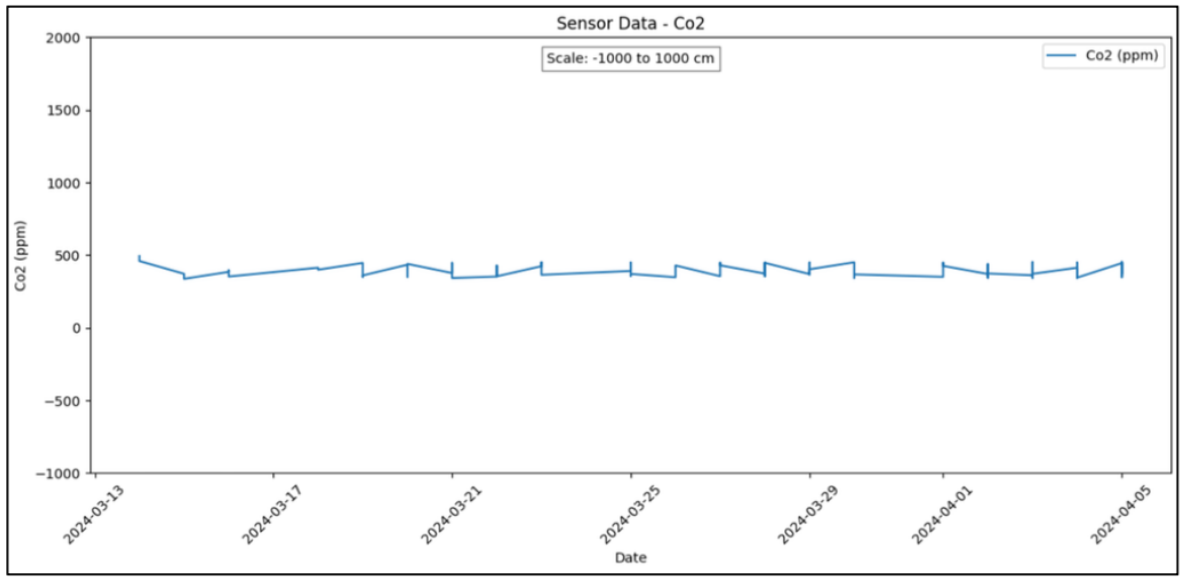

*Figure_13: Data collected from CO2 sensor*

- **TDS (Total Dissolved Solids):** The TDS sensor was used to detect levels of nutrients in the water. The ideal TDS range varies depending on the plant species, growth stage, and nutrient solution formulation. Generally, TDS in the range between 800 and 1500 parts per million (ppm) is suitable for most hydroponic crops. The data received from the TDS sensor is visualized in the *Figure_14*, below:

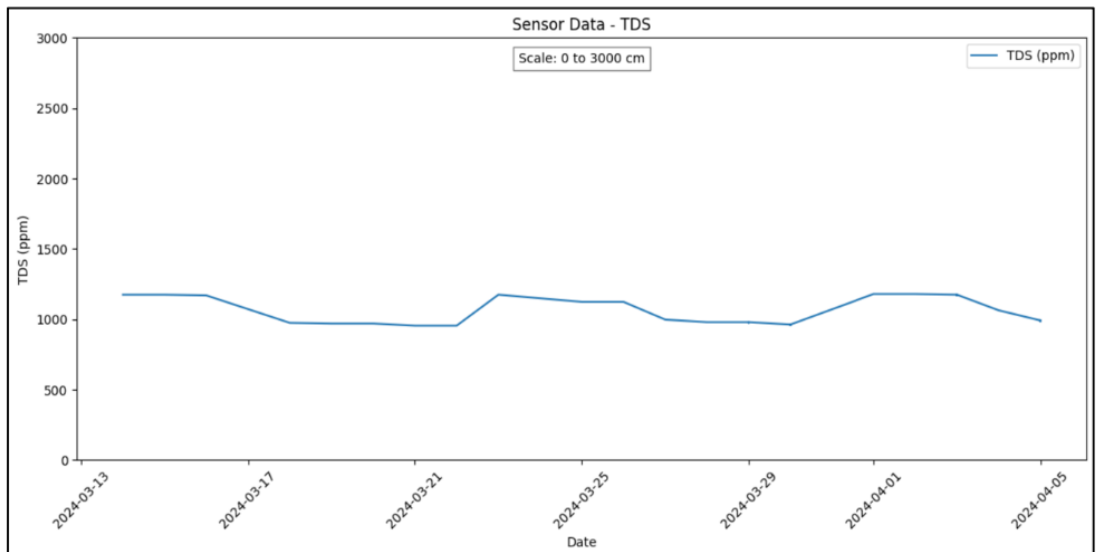

*Figure_14: Data collected from the TDS sensor*

- **TEMP:** The temperature sensor was used to measure the thermal conditions surrounding the plants. Low temperatures are typically values below 15°C. Medium temperature ranges, considered favorable for most plants, are between 15°C and 25°C, which provides an ideal balance for metabolic processes and growth. High temperature levels, exceeding 25°C, would cause an excess of heat that could lead to stress and reduced growth rates of plants. The data received from the temperature sensor is visualized in the *Figure_15*, below:

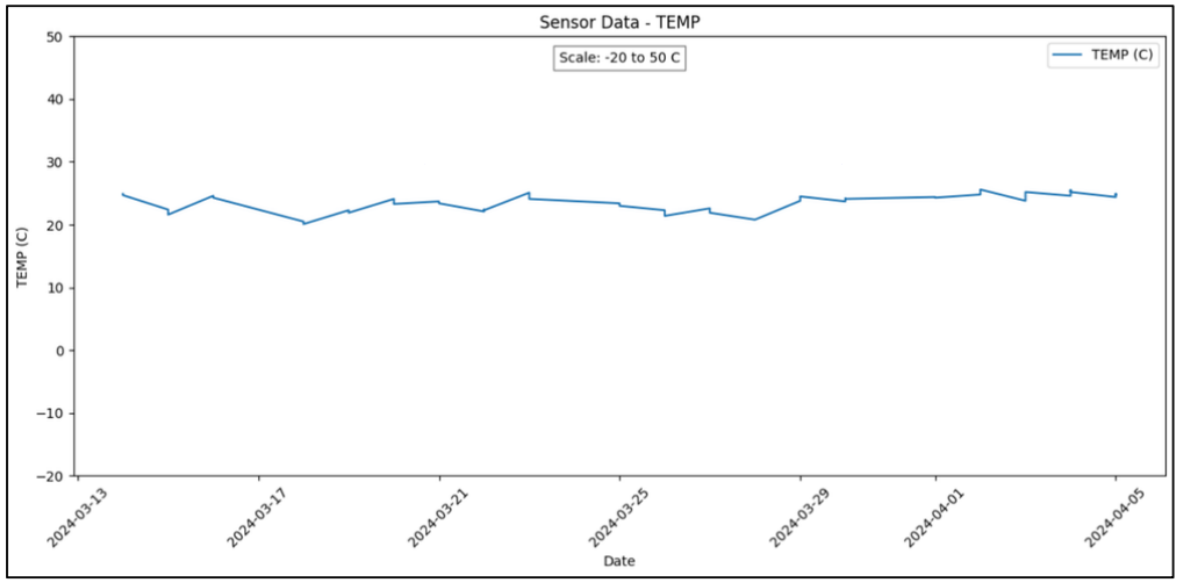

*Figure_15: Data collected from the temperature sensor*

- **HUM:** The humidity sensor was used to measure the moisture of the air surrounding the plants. Low humidity levels are often considered to be below 40% relative humidity (RH). Medium humidity levels are between 40% and 60% RH, and are generally good for most plants, providing sufficient moisture for transpiration and nutrient uptake while avoiding the risk of water stress or dehydration. High humidity levels, exceeding 60% RH, would cause excess moisture in the air that could lead to fungal diseases, reduced nutrient uptake, and reduced plant growth. The data received from the humidity sensor is visualized in the *Figure_16*, below:

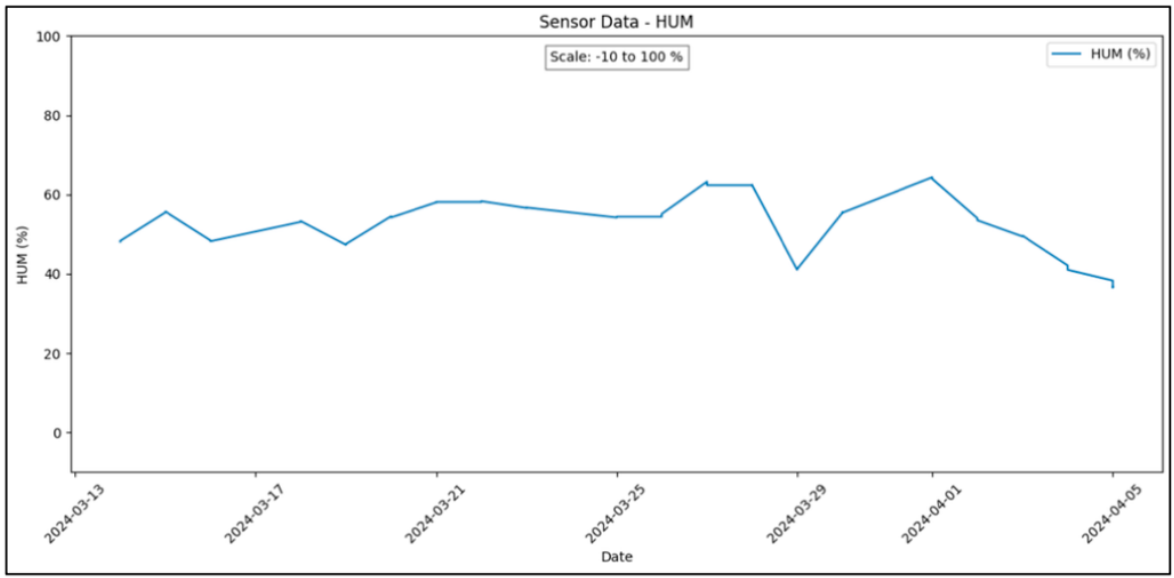

*Figure_16: Data collected from the humidity sensor*

- **WaterTemp:** The water temperature was used to measure the thermal conditions of the water under the plant roots. Low water temperatures are usually below 10°C. Medium water temperatures are between 10°C and 20°C and are considered suitable for most plants. High water temperatures, exceeding 20°C, may cause an excess of heat in the water which could lead to many issues like reduced dissolved oxygen levels or root rot. The data received from the water temperature sensor is visualized in the figure_17, below:

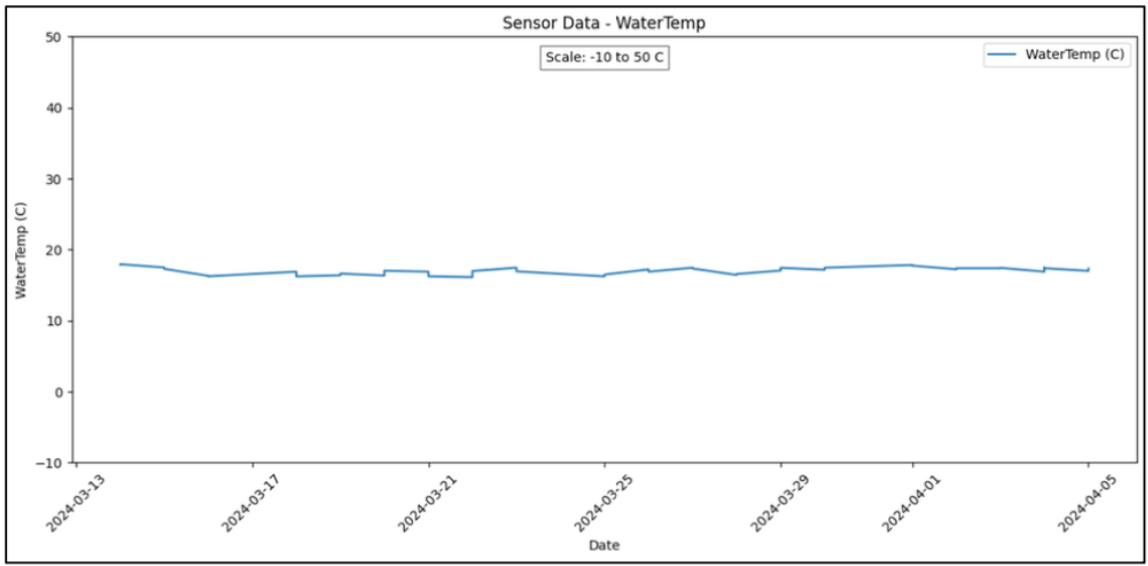

*Figure_17: Data collected from the water temperature sensor*